\documentclass[twocolumn,aps,pra,superscriptaddress]{revtex4}
\usepackage{graphicx,amsmath,amssymb}
\usepackage{dcolumn,fancyhdr}
\usepackage{bm,natbib,mathrsfs}
\usepackage{times}
\usepackage{txfonts}
\usepackage{epstopdf}
\usepackage{latexsym}
\usepackage{epsfig,palatino}
\usepackage{color}

\newcommand{\ket}[1]{\left\vert#1\right\rangle}

\newcommand{\bra}[1]{\left\langle#1\right\vert}

\newcommand{\nbar}{\overline{n}}

\begin{document}

\title{Out of equilibrium thermodynamics of quantum harmonic chains}

\author{A. Carlisle} 
\affiliation{Centre for Theoretical Atomic, Molecular and Optical Physics, School of Mathematics and Physics, Queen's University, Belfast BT7 1NN, United Kingdom}
\author{L. Mazzola}
\affiliation{Centre for Theoretical Atomic, Molecular and Optical Physics, School of Mathematics and Physics, Queen's University, Belfast BT7 1NN, United Kingdom}
\author{M. Campisi}
\affiliation{Institute of Physics, University of Augsburg, D-86153 Augsburg, Germany}
\author{J. Goold}
\affiliation{The Abdus Salam International Centre for Theoretical Physics, 34014 Trieste, Italy}
\author{ F. L. Semi\~ao}
\affiliation{Centro de Ci\^encias Naturais e Humanas, Universidade Federal do ABC, 09210-170, Santo Andr\'e, S\~ao Paulo, Brazil}
\author{ A. Ferraro}
\affiliation{Centre for Theoretical Atomic, Molecular and Optical Physics, School of Mathematics and Physics, Queen's University, Belfast BT7 1NN, United Kingdom}
\author{F. Plastina}
\affiliation{Dipartimento di Fisica \& INFN--Gruppo collegato di Cosenza, Universit\`a della Calabria,
       Via P.Bucci, 87036 Arcavacata di Rende (CS), Italy}
\author{V. Vedral}
\affiliation{Department of Physics, University of Oxford, Clarendon Laboratory, Oxford, OX1 3PU, UK}
\affiliation{Center for Quantum Technology, National University of Singapore, Singapore}
\author{ G. De Chiara}
\affiliation{Centre for Theoretical Atomic, Molecular and Optical Physics, School of Mathematics and Physics, Queen's University, Belfast BT7 1NN, United Kingdom}
\author{M. Paternostro}
\affiliation{Centre for Theoretical Atomic, Molecular and Optical Physics, School of Mathematics and Physics, Queen's University, Belfast BT7 1NN, United Kingdom}
\affiliation{Institut f\"ur Theoretische Physik, Albert-Einstein-Allee 11, Universit\"at Ulm, D-89069 Ulm, Germany}
\date{\today}

\begin{abstract}
The thermodynamic implications for the out-of-equilibrium dynamics of quantum systems are to date largely unexplored, especially for quantum many-body systems. In this paper we investigate the paradigmatic case of an array of nearest-neighbor coupled quantum harmonic oscillators interacting with a thermal bath and subjected to a quench of the inter-oscillator coupling strength. We study the work done on the system and its irreversible counterpart, and characterize analytically the fluctuation relations of the ensuing out-of-equilibrium dynamics. Finally, we showcase an interesting functional link between the dissipated work produced across a two-element chain and their degree of general quantum correlations. Our results suggest that, for the specific model at hand, the non-classical features of a harmonic system can influence significantly its thermodynamics.
\end{abstract}

\maketitle

The out-of-equilibrium dynamics of quantum systems offer a very interesting stage for the study of the thermodynamic properties~\cite{esposito,campisi,seifert}. The establishment of quantum fluctuation theorems represents a milestone in the link between arbitrarily fast quantum dynamics and equilibrium figures of merit of thermodynamic relevance, such as feee energy changes, heat, work, and entropy~\cite{Tasaki,Crooks,Jarzynski}. The definition of such quantities from a genuine quantum mechanical standpoint, the formulation of their operational interpretations, and the design of experimental techniques for their quantitative assessment are some of the drives of current research on the thermodynamic properties of quantum systems and processes~\cite{lutz,varie,Abah,Dorner,Joshi,Huber,Heyl,Ngo,CPmaps,Oxf,KavanJohn,JohnMauroKavan,Batalhao}. An extensive programme of investigations aimed at understanding and characterising the non-equilibrium thermodynamics of simple, paradigmatic systems is currently underway, including exactly solvable extended spin models~\cite{Joshi,Dorner,Smacchia,Joshi13EPJB86,Sindona,dudu,Fusco}, which have offered an interesting platform for the study of the emergence of irreversible thermodynamics from quantum many-body features~\cite{Joshi,Dorner}.

In this context, a rather privileged role is played by the quantum oscillator, which offers the possibility for the (either exact or approximate) analytical assessment of non equilibrium features in an ample range of situations, including external driving and special nonlinear cases~\cite{Galve,varieoscillator}. However, to the best of our knowledge, little is known on composite systems consisting of more than a single harmonic oscillator. This is an interesting case to study, as it would enable the assessment of the scaling properties of thermodynamically relevant quantities with the size of the system, as well as the study of processes involving either the whole system or only part of it, which in principle would result in different behaviors and manifestations.

This is precisely the context within which the investigation reported in this paper lies. We aim at addressing the effects that a global quench of the inter-particle coupling strength has on the phenomenology of thermodynamic quantities such as (irreversible) work and free energy differences. We study the case of an open-ended array of quadratically coupled quantum harmonic oscillators, in contact with a thermal reservoir. By allowing for a global quantum quench, we address the scaling of both the average work and the free energy differences, providing exact analytic expressions for the dissipated work, which is an important figure of merit to gauge the deviations of the actual state of the array after the quench from its counterpart at thermodynamic equilibrium. It thus gives us information about the effects of non-adiabaticity. However, this study offers even more opportunities for exploration: by calculating  explicitly the amount of quantum correlations shared by the elements of a two-oscillator system, we illustrate the existence of a clear functional relation between dissipated work and quantum correlations. For the specific case of the coupling model at hand, this hints at the interdependence of quantum and thermodynamic features in quadratically coupled harmonic chains. This is a tantalising possibility that will deserve future in-depth explorations. 

The remainder of this paper is organised as follows: Sec.~\ref{modello} introduces the harmonic model and illustrates an interferometric approach to the exact determination of the characteristic function of work distribution~\cite{campisi} resulting from a sudden quench of the inter-oscillator coupling strength. This opens the way to the assessment of quantum fluctuation relations~\cite{Tasaki,Crooks,Jarzynski} and the fully analytic calculation of the average work, free energy change and other figures of merit for the characterization of irreversibility. This study allows us to identify the degree of squeezing generated by the oscillators' quadratic coupling as a very important resource for the ability of the process to do work on the system (see Ref.~\cite{Galve} for a different analysis of this point made on a single harmonic oscillator). Our calculations, which are valid for chains of an arbitrary number of elements, allow for the clear identification of ``classical" and ``quantum" parts of both the change of free energy, which are then related to the degree of quantum correlations across a two-element chain in Sec.~\ref{correlations}. Finally, in Sec.~\ref{conc} we draw our conclusions and discuss briefly the questions opened by our study. Two appendices summarize the most technical part of our calculations. 

\section{Description of the coupling model and analysis of nonequilibrium thermodynamics} 
\label{modello}

\begin{figure*}[t]
\hskip1.3cm{\bf (a)}\hskip7.8cm\centering{\bf (b)}\\
\includegraphics[width=0.5\linewidth]{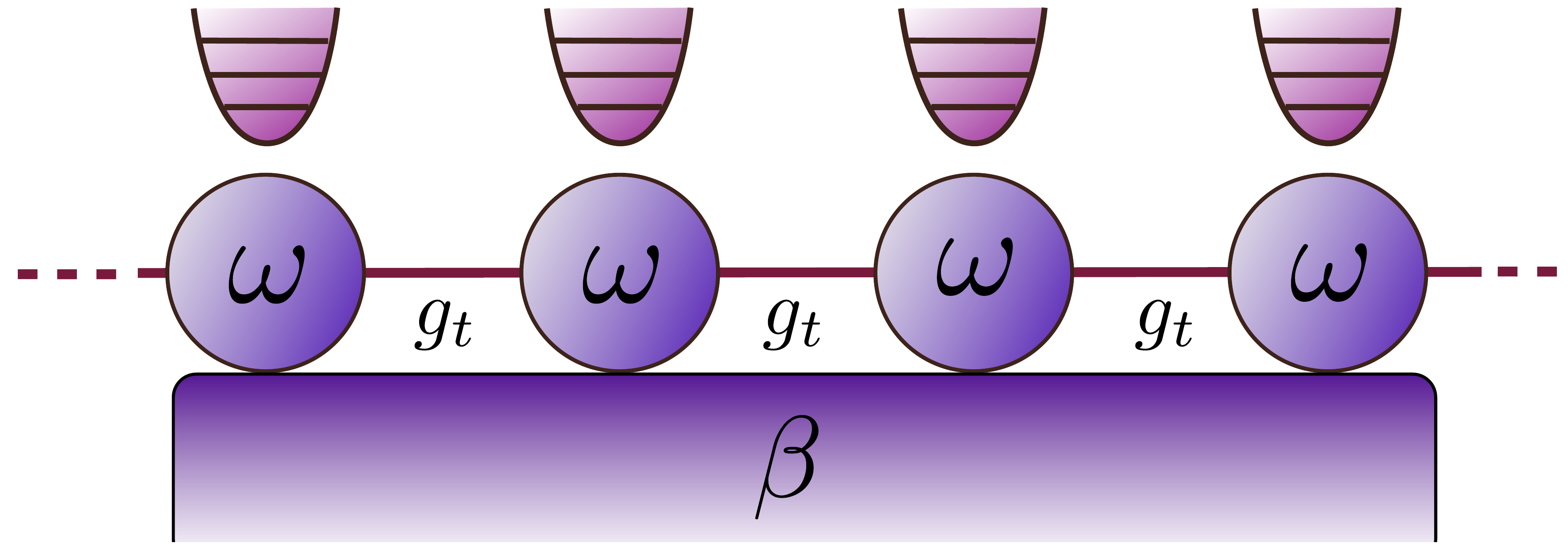}~~~~\includegraphics[width=0.8\columnwidth]{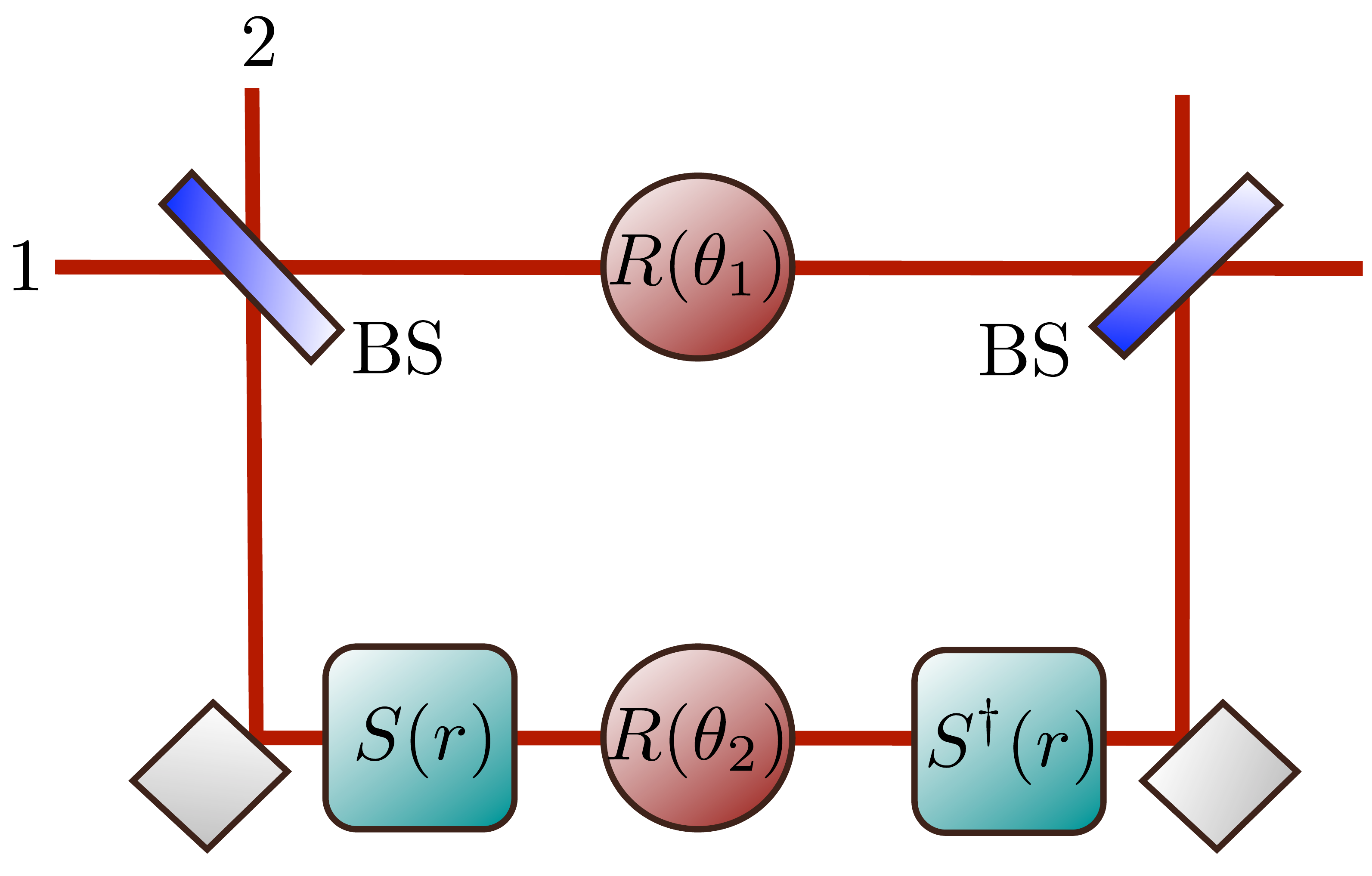}\\
\caption{(Color online) {\bf (a)} Sketch of the model considered in this paper: a linear chain of coupled harmonic oscillators (coupling strength $g_t$) is in contact with a thermostat at inverse temperature $\beta$. The couplings are all suddenly quenched to bring the system out-of-equilbrium. We study the thermodynamics of the corresponding evolution. {\bf (b)} Equivalent interferometer describing, in terms of linear optics elements, the time evolution resulting from the propagator $e^{-i\hat{\cal H}_1t}$ generated by the quenched model in Eq.~\eqref{model}. We show the symbols for single-mode squeezing [$S(r)$], phase-space rotation [$R(\theta)$], and two-mode beam splitting [BS].} 
\label{equivalente} 
\end{figure*}

We consider coupled harmonic oscillators in an open linear configuration [cf. Fig.~\ref{equivalente} {\bf (a)}]. While in this part of our analysis we will mostly concentrate on the case of only two coupled oscillators, the generalization to a multi-element register is addressed later on. We start from a Hooke-like coupling model between two harmonic oscillators in contact with a heat bath at temperature $T$. The model is described by the following Hamiltonian (we assume units such that $\hbar=1$ across the manuscript)
\begin{equation}
\label{model}
\hat{\cal H}_1(g_t)=\frac{\Omega}{2}\sum^2_{j=1}(\hat x^2_j+\hat p^2_{j})+g_t(\hat x_1-\hat x_2)^2
\end{equation}
with $\Omega$ the frequency of the oscillators (assumed for simplicity to be identical and with a unit mass) and $g$ the (possibly) time-dependent interaction strength. Here, $\hat x_j$ and $\hat p_j$ are the position- and momentum-like operators of oscillator $j=1,2$ (satisfying the commutation relations $[\hat x_j,\hat p_j]=i$). Within the context of our analysis we will assume that, after detaching the system from the heat bath, the coupling strength is abruptly turned on to the value $g_0>0$, namely $g_t=g_0\Theta(t)$, where $\Theta(t)$ is the Heaviside step function. This process embodies a sudden quench of the interaction between the harmonic oscillators. A straightforward calculation shows that the post-quench time evolution operator $\hat{\cal U}(t>0)=e^{-i\hat{\cal H}_1t}$ generated by Eq.~\eqref{model} can be written as 
\begin{equation}
\label{deco}
\hat{\cal U}(t>0)=\hat{\cal B}^\dag\hat{\cal S}^\dag(r)[\hat{\cal R}_1(\theta_1(t))\otimes\hat{\cal R}_2(\theta_2(t))]\hat{\cal S}(r)\hat{\cal B},
\end{equation}
where $\hat{\cal B}=\exp[\pi(\hat x_1\hat p_2-\hat x_2\hat p_1)/4]$ is the $50:50$ beam-splitter operator, $\hat{\cal S}(r)=\hat\openone_1\otimes\hat{\cal S}_2(r)$ describes the local squeezing of oscillator $2$ by a degree $r=(1/4)\ln\!\sqrt{1+2g_0/\omega}$ performed by the squeezing operator $\hat{\cal S}_2(r)=\exp[i{\rm Im}(r)(\hat x^2_2-\hat p^2_2)-i{\rm Re}(r)(\hat x_2\hat p_2+\hat p_2\hat x_2)]$
, $\omega=\Omega/2$ and $\hat{\cal R}_j(\theta_j)=\exp[-i\theta_j(\hat{x}^2_j+\hat{p}^2_j)]$ accounts for phase-space rotations by the angle $\theta_j$~($j=1,2$). In the specific case of our problem we have $\theta_1(t)=\omega t$ and $\theta_2(t)=\omega t\sqrt{1+2g_0/\omega}$. In light of such decomposition, which accounts for the free evolution (each occurring at the respective frequency) of the centre-of-mass and relative-motion modes of the system, the time-evolution of the two-oscillator system can be understood as the result of the action of a Mach-Zehnder interferometer endowed with an {\it active} element, embodied by the local squeezer, on one of its arms [cf. Fig.~\ref{equivalente}(b)]. This establishes quantum correlations between the harmonic oscillators. Our first goal here is to show that such correlations are linked with the work that is irreversibly generated in the process due to the non-adiabatic nature of the quench. 

\begin{figure}[!b]
\centering{\bf (a)}\hskip3cm{\bf (b)}\\
\includegraphics[width=\linewidth]{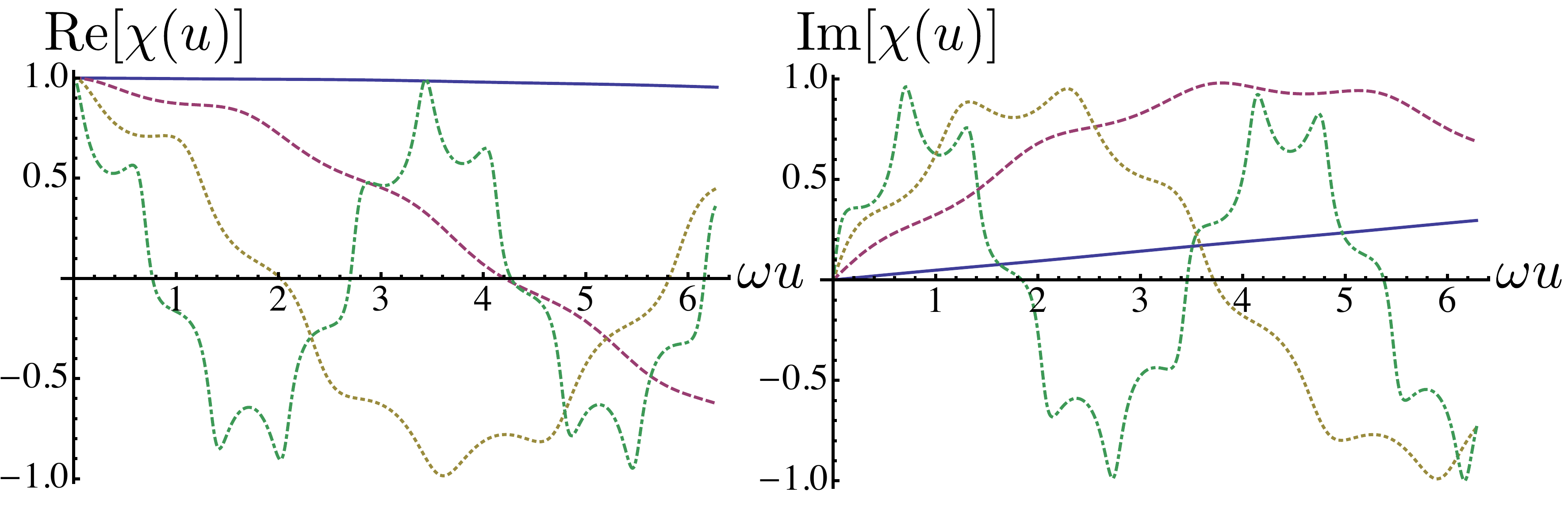}\\
\centering{\bf (c)}\hskip3cm{\bf (d)}\\
\includegraphics[width=\linewidth]{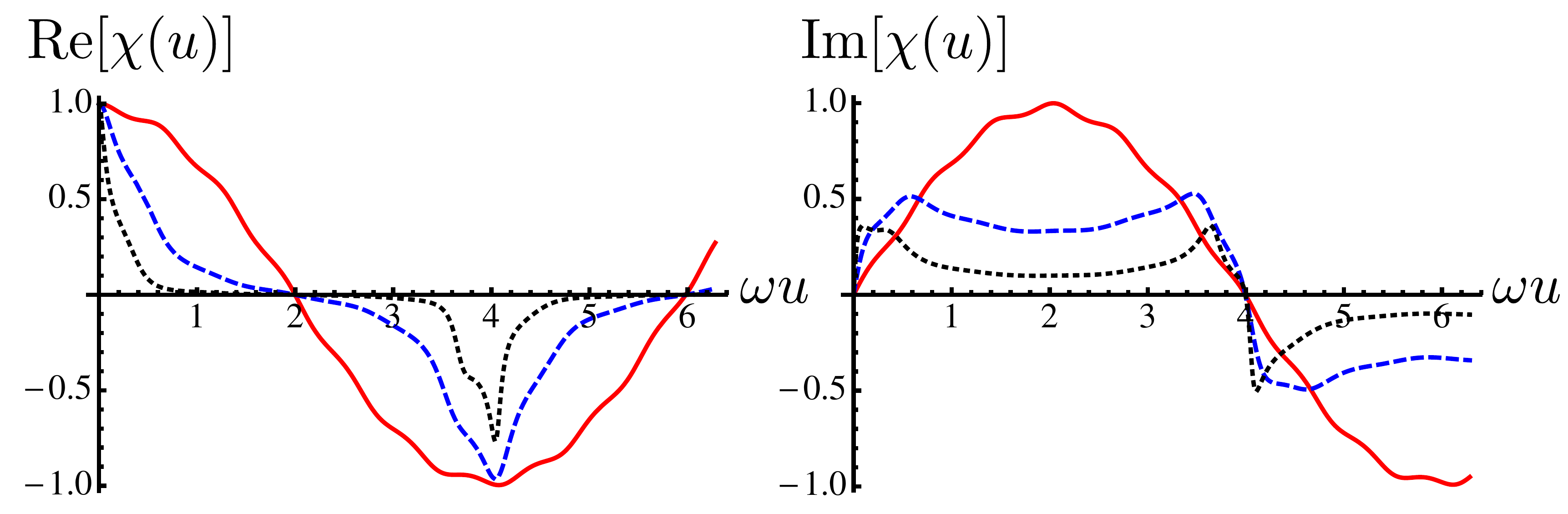}
\caption{(Color online) Panels {\bf (a)} and {\bf (b)}: Characteristic function of the work distribution after a sudden quench of the coupling strength between two harmonic oscillators coupled via a Hooke-like model. We show the behavior of ${\rm Re}[\chi(u)]$ [panel {\bf (a)}] and ${\rm Im}[\chi(u)]$ [panel {\bf (b)}] against $\omega{u}$ for $V=1$ and $g_0/\omega=0.1$ (solid line), $1$ (dashed line), $3$ (dotted line), and $10$ (dot-dashed line). Panels {\bf (c)} and {\bf (d)}: Same study as in panels {\bf (a)} and {\bf (b)} but for $g_0/\omega=0.75$ and $V=1$ (solid line), $3$ (dashed line), and $10$ (dotted one).} 
\label{charfunc} 
\end{figure}

In order to accomplish this goal, let us briefly sketch the way to compute the characteristic function of the work probability distribution associated with the process that takes abruptly the Hamiltonian from $\hat {\cal H}_i\equiv\hat{\cal H}_1(0)$ to $\hat{\cal H}_f=\hat{\cal H}_1(g_0)$ at time $t=0$. As we will show, $\chi(u)$ can be understood in terms of the thermal convolution of inner products between displaced squeezed vacuum states. For the sudden switch of the work parameter that we are considering here, the expression for the characteristic function of work distribution takes the form
\begin{equation}
\chi(u)={\rm Tr}[e^{iu\hat{\cal H}_f}e^{-iu\hat{\cal H}_i}\rho^{th}_S(0)],
\end{equation}
where $\rho^{th}_S(0)=e^{-\beta\hat{\cal H}_i}/{\cal Z}_0$ is a pre-quench thermal-equilibrium state of the two harmonic oscillators at inverse temperature $\beta$ and ${\cal Z}_0={\rm Tr}[e^{-\beta\hat{\cal H}_1(0)}]$ is the associated partition function. In light of the structure shown in Eq.~\eqref{deco}, it is convenient to decompose the pre-quench state over the single-oscillator coherent-state basis as $\rho^{th}_S(0)=\int\,d^2\alpha_1\,d^2\alpha_2\prod^2_{j=1}P^{th}_{V}(\alpha_j)\ket{\alpha_1,\alpha_2}\bra{\alpha_1,\alpha_2}_{12}$ with $P^{th}_{V}(\alpha_j)=2[\pi(V-1)]^{-1}\exp[{-{2|\alpha^2_j|}/({V-1})}]$ the thermal $P$-function of oscillator $j$, characterised by the variance $V=2\nbar+1$ with $\nbar=(e^{\beta\omega}-1)^{-1}$ the thermal mean occupation number. Here, $\vert{\alpha_j}\rangle=\hat{\cal D}_j(\alpha_j)\ket{0}_j$ is a coherent state generated by the displacement operator $\hat{\cal D}_j(\alpha_j)=\exp[\alpha_j\hat a^\dag_j-\alpha^*_j\hat a_j]$ over the vacuum. With this at hand, we have
\begin{equation}
\label{total}
\chi(u)=\int\!d^2\alpha_1\,d^2\alpha_2\prod^2_{j=1}P^{th}_{V}(\alpha_j)\,\chi_{\alpha_1,\alpha_2}(u)
\end{equation}
with $\chi_{\alpha_1,\alpha_2}(u)=
\langle{\alpha_1,\alpha_2}\vert e^{i\hat{\cal H}_f u}e^{-i\hat{\cal H}_i u}\ket{\alpha_1,\alpha_2}$ the Loschmidt echo corresponding to the evolution of a pair of initial coherent states under the process addressed here. As the interaction between the harmonic oscillators is quadratic, the Gaussian nature of coherent states is preserved across the process, and the thermal convolution in Eq.~\eqref{total} consists  of a four-fold integration over Gaussian functions. We thus focus on the explicit evaluation of $\chi_{\alpha_1,\alpha_2}(u)$, whose details are given in the Appendix, and results in the elegant expression $\chi_{\alpha_1,\alpha_2}(u)=\!\bra{\zeta_1;\xi_1}{\zeta_2;\xi_2}\rangle$ with $|{\zeta_j;\xi_j}\rangle=\hat{\cal D}_j(\zeta)\hat{\cal S}_j(\xi)\ket{0}_j$ a displaced squeezed state~($\zeta,\xi\in\mathbb{C}$)~\cite{Caves,Moeller}, which can be calculated analytically to be
\begin{widetext}
\begin{equation}
\chi_{\alpha_1,\alpha_2}(u)=\frac{\exp\left[{\dfrac{[({\zeta_2}-{\zeta_1})\sinh{r}+({\zeta^*_1}-{\zeta^*_2})\cosh{r}][ ({\zeta_2}-{\zeta_1})\cosh{r}+e^{2 i \theta_2 (u)}({\zeta^*_1}-{\zeta^*_2})\sinh{r}]}{2(\cosh ^2r-\sinh^2r\, e^{2 i \theta_2 (u)})}-\dfrac{\zeta_1\zeta^*_2-\zeta^*_1\zeta_2}{2}}\right]}{\sqrt{\cosh^2r-e^{2 i \theta_2 (u)}\sinh^2{r}}}.
\end{equation}
The expressions for $\zeta_{1,2}$ and $\xi_{1,2}$ are given in the Appendix. Examples of the behavior of the characteristic function for various quench strengths $g_0$ and temperatures of the initial equilibrium states  are shown in Fig.~\ref{charfunc}.
\end{widetext}

Looking at Fig.~\ref{charfunc} ${\bf (c)}$ and ${\bf (d)}$, we see that as the temperature of the initial thermal states increases ({\it i.e.}, as $V$ grows), the absolute value of the derivative of both the real and the imaginary part of $\chi(u)$ at $u=0$ grows. This is an important observation in light of the possibility to evaluate the average work extractable from the system after the process as $\langle W\rangle=-i\partial_u\chi(u)\vert_{u=0}$. Although the full-fledged expression of $\chi(u)$ at arbitrary values of $\beta$ is too involved to be reported here, the average work takes the compact expression $\langle W\rangle=g_0 V/2$, which is thus linear in the strength of the quench and takes the frequency-independent value $g_0/2$ in the low temperature limit $\beta\to\infty$ and grows as $g_0/(\beta\omega)$ in the classical limit for very large temperatures.

As a check that our analytic form for the characteristic function is correct we consider the Jarzynski equality $\chi(i\beta)=e^{-\beta\Delta{F}}$. The net change in free energy of the system can be evaluated using the pre- and post-quench partition functions ${\cal Z}_0$ and ${\cal Z}$, whose evaluation we now sketch. While the calculation of the pre-quenched case trivially leads to ${\cal Z}_0=4/\sinh^2(\beta\omega/2)$, in line with the tensor-product nature of the initial equilibrium state, the post-quenched one requires the evaluation of 
\begin{equation}
\begin{aligned}
{\cal Z}&={\rm Tr}[e^{-\beta\hat{\cal H}(g_0)}]={\rm Tr}[\hat{\cal B}^\dag\hat{\cal S}^\dag e^{-\sum^2_{j=1}\theta_j(\beta)(\hat x^2_j+\hat p^2_j)}\hat{\cal S}\hat{\cal B}]\\
&={\rm Tr}[e^{-\sum^2_{j=1}\theta_j(\beta)(\hat x^2_j+\hat p^2_j)}]=\frac{4}{\sinh(\beta\omega/2)\sinh(\theta_2(\beta)/2)}
\end{aligned}
\end{equation}
so that $e^{-\beta\Delta{F}}=\sinh\left(\frac{\beta\omega}2\right){\rm csch}\left(\frac{\beta\omega}{2}\sqrt{1+\frac{2g_0}{\omega}}\right)$. This in turn gives us the free-energy change
\begin{equation}
\Delta{F}=-\frac{1}{\beta}\ln\left[\frac{\sinh(\beta\omega/2)}{\sinh\left(\frac{\beta\omega}{2}\sqrt{1+2g_0/\omega}\right)}\right].
\end{equation}
In the classical limit of very high temperature, this expression becomes $\Delta{F}_c\simeq(1/\beta)\ln[\sqrt{1+2g_0/\omega}]$. In the quantum limit of $\beta\to\infty$, on the other hand, the net change in free energy is bound by the asymptotic value $\Delta{F}_{q}\simeq(\omega/2)(\sqrt{1+2g_0/\omega}-1)$, which only depends on the strength of the quench (in units of $\omega$). Although we have not been able to study analytically the Jarzynski identity due to the cumbersome form of $\chi(u)$, we have numerically checked that it is satisfied.

\begin{figure*}[t]
{\centering {\bf (a)}\hskip5cm{\bf (b)}\hskip6cm{\bf (c)}}
\includegraphics[width=0.7\columnwidth]{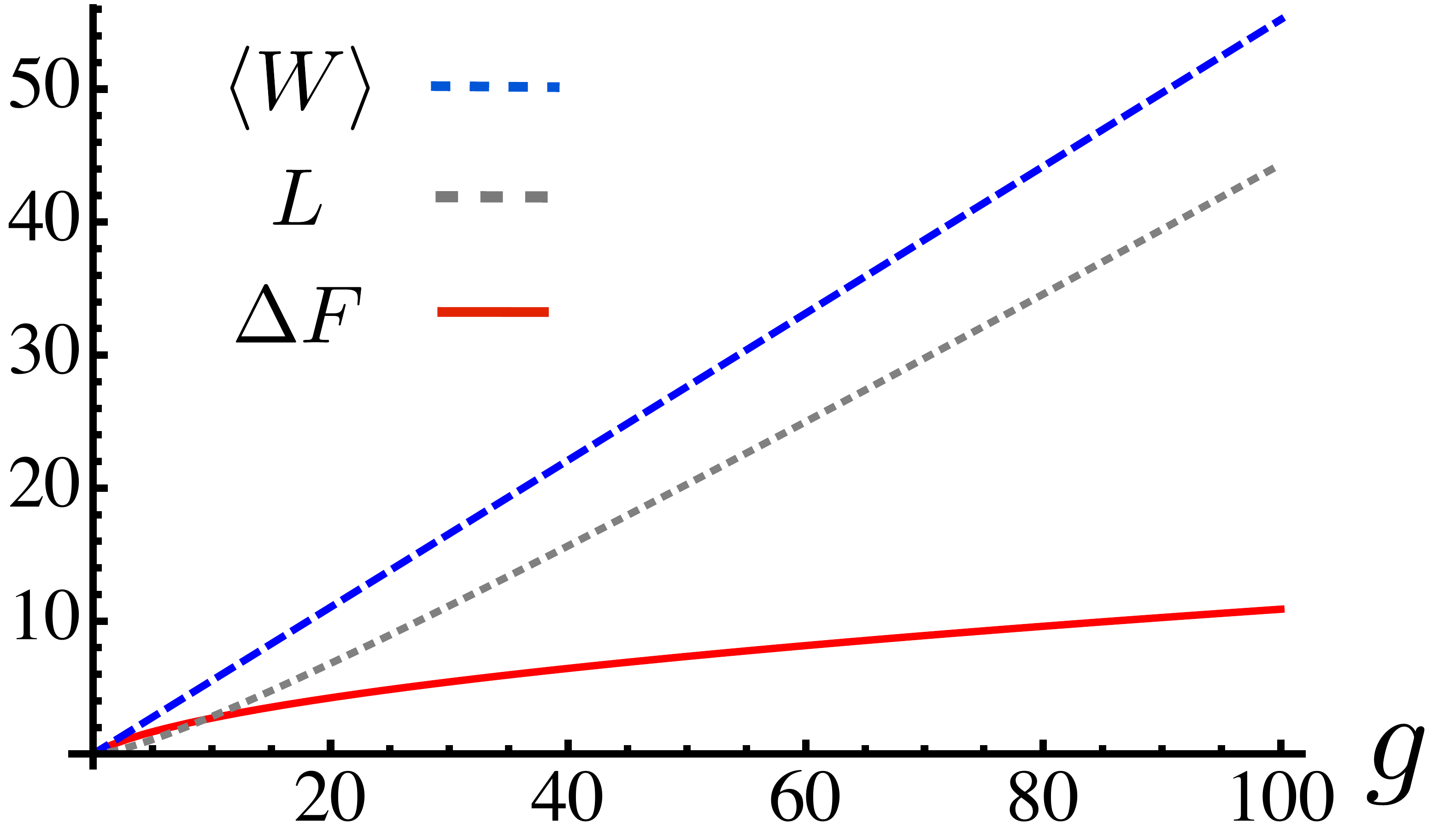}\includegraphics[width=0.7\columnwidth]{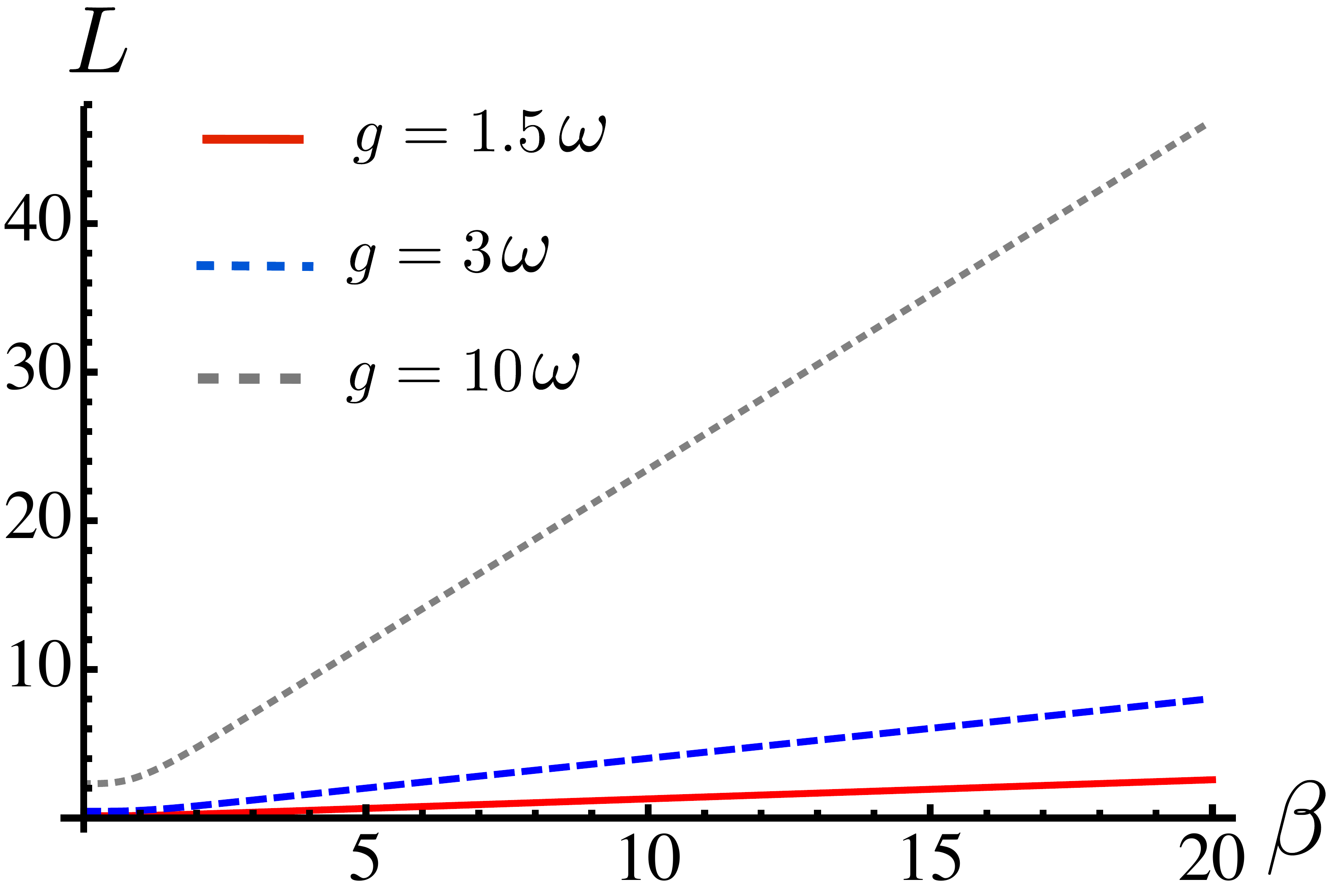}\includegraphics[width=0.7\columnwidth]{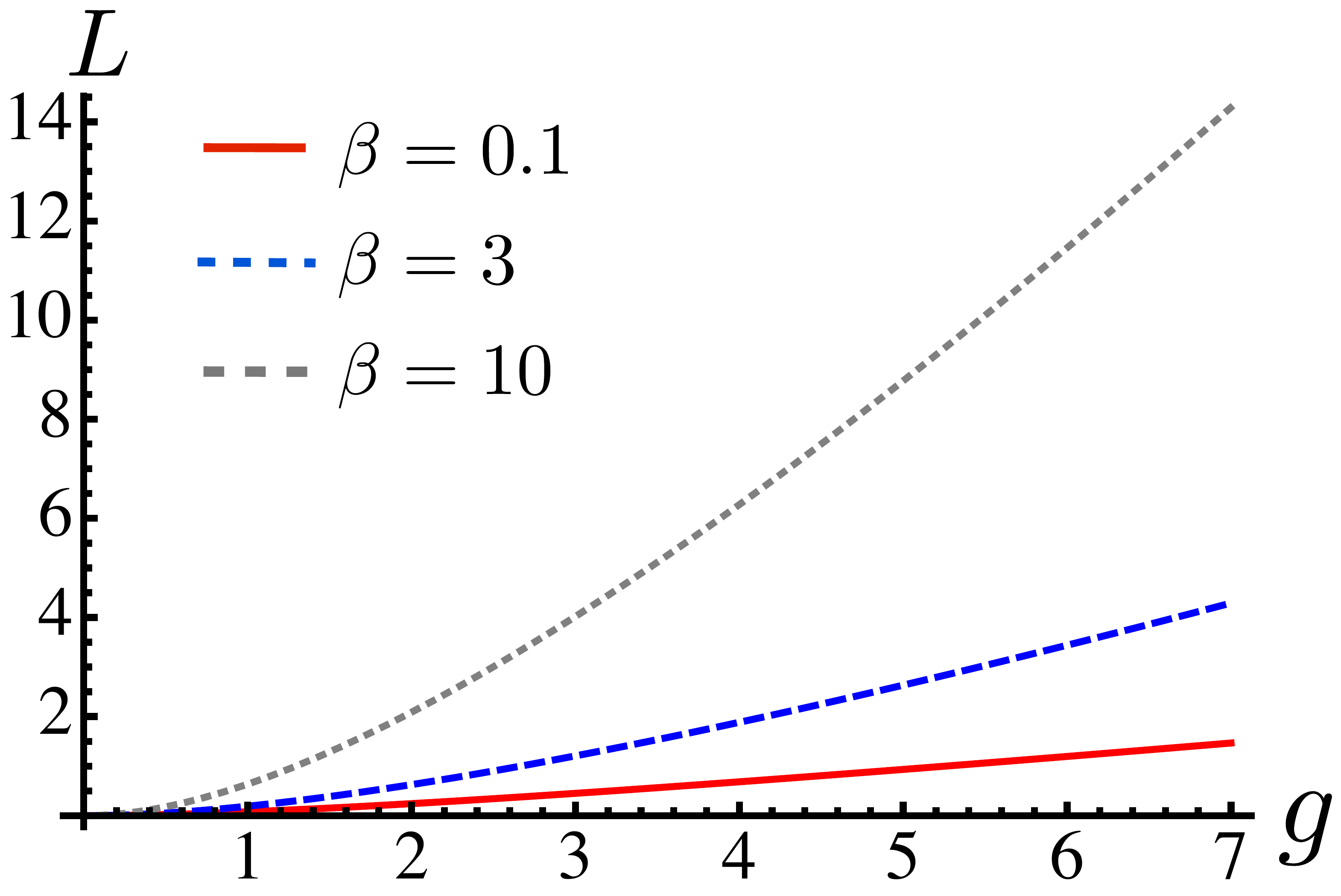}
\caption{(Color online) {\bf (a)} We plot the average work $\langle W\rangle$, the free-energy change $\Delta F$ and the correspondingly produced nonequilibrium lag for a system of two oscillators with $\omega=3$ and $\beta=1$ against the coupling strength $g$. {\bf (b)} [{\bf (c)}] We study of the nonequilibrium lag produced for the system addressed in panel {\bf (a)} against the inverse temperature [coupling strength], for three different values of the strength of the quench [three values of the inverse temperature].} 
\label{studio} 
\end{figure*}

We now analyze the degree of irreversibility of our quench  process. This can be quantified by
the quantity
\begin{equation}
L = \beta W_\text{diss} = \beta [\langle W\rangle-\Delta{F}] = D[\rho_t||\rho_t^{eq}],
\end{equation}
which accounts for the ``nonequilibrium lag'' between the actual system state $\rho_t$ and the reference
thermal state $\rho_t^{eq}=e^{-\beta \hat{\cal H}(t)}/{\cal Z}(t)$
as measured by the Kullback-Leibler divergence (or relative entropy) between two arbitrary states $\rho$ and $\sigma$ and defined as $D[\rho ||\sigma]=\text{Tr}(\rho\log\rho-\rho\log\sigma)$~\cite{Bochkov81aPHYSA106,Schloegl66ZP191,Vaikuntanathan09EPL87,Deffner10PRL105}. We find 
\begin{equation}
L=\frac{\beta  g_0}{2} \coth \left(\frac{\beta \omega}{2}\right)+\ln \left[\sinh \left(\frac{\beta\omega }{2}\right) \text{csch}\left(\frac{\beta\omega}{2}  \sqrt{\frac{2
   g_0}{\omega}+1}\right)\right].
\end{equation}
Despite being customarily referred to as ``nonequilibrium entropy production'', $L$ is in general not equal to 
the change in thermodynamic entropy~\cite{Joshi13EPJB86}, hence we dub it more appropriately  the ``nonequilibrium lag''. In Fig.~\ref{studio} we report the analysis of average work, change in free energy, and nonequilbrium lag against the strength of the quench, as well as the assessment of the dependence of $L$ on the inverse temperature and $g$. A remarkable feature is the quasi-linear growth of the nonequilibrium lag at low temperatures [cf. Fig.~\ref{studio} {\bf (b)}], which will be useful for the analysis reported in Sec.~\ref{correlations}.

Another closely related quantifier of irreversibility, specifically designed for thermally  isolated systems,
is provided by 
\begin{equation}
\Delta{\cal E} = {\rm Tr }\, [ \rho_t \hat{\cal E} (t)-  \rho_0 \hat{\cal E}(0)],
\end{equation}
which is defined using the operator
\begin{equation}
\hat{\cal E}(t) = \sum_k \ln k |k,t\rangle \langle k,t|
\end{equation}
built using the eigenstates  $|k,t\rangle$ of the instantaneous Hamiltonian $H(t)$. They are ordered by their
increasing energy $E_k(t)> E_m(t)$ for $k>m$. 
The operator $\hat{\cal E}$, first introduced in Ref.~\cite{Michele},
is the quantum version of the Gibbs entropy associated with the microcanonical ensemble
\cite{GibbsBook,Hertz10AP338a,Einstein11AP34,BeckerBook,Campisi08PRE78b,MuensterBook,Campisi05SHPMP36,Dunkel14NATPHYS10}. 
Just like thermodynamic entropy, it remains unchanged in a slow (adiabatic) protocol and
cannot decrease in a generic fast one, provided the initial density matrix is diagonal in the initial Hamiltonian eigenbasis, its eigenvalues are ordered in a non-increasing fashion, and the spectrum is non-degenerate at all times. The quantitative analysis of the behavior of $\Delta{\cal E}$ in our system, which is made possible by the knowledge of the spectrum of 
$\hat{\cal H}_1$ as obtained in the Appendix, will be presented elsewhere~\cite{tocome}.

We now turn to the assessment of the role that squeezing has on the ability of the system to produce extractable work. In order to do so, we compare the performance of the coupling scheme addressed so far to the ability of the system to perform work when the two harmonic oscillators are coupled via the model $\hat x_1\hat p_2-\hat p_1\hat x_2$ That is, we consider the Hamiltonian
\begin{equation}
\begin{aligned}
\hat{\cal H}_2&=\frac{\Omega}{2}\sum^2_{j=1}(\hat x^2_j+\hat p^2_j)+g_t(\hat x_1\hat p_2-\hat p_1\hat x_2).
\end{aligned}
\end{equation}
There are two fundamental differences between $\hat{\cal H}_1$ and $\hat{\cal H}_2$: first, $\hat{\cal H}_2$ is energy preserving and the corresponding time propagator would not require the squeezing of any harmonic oscillator~\cite{Helen}. 
As we will argue soon, this gives rise to key differences with respect to the thermodynamic behavior showcased up to this point. Second, consistently with the fact that $\hat{\cal H}_2$ is the rotating-wave form of Eq.~\eqref{model}, the strength of the quench cannot be arbitrary, as the spectrum of the Hamiltonian acquires an imaginary eigenvalue for $g_0>\Omega$.

Besides this limitation, the characteristic function associated with the process generated by a quench of $\hat{\cal H}_2$ can be worked out in a way similar to what has been sketched before for the case of Eq.~\eqref{model}. A second-order Taylor expansion of the characteristic function with respect to variable $u$ leads to the approximate expression 
$\chi_{\hat{\cal H}_2}(u)\simeq1-\frac{g^2_0}{16}(V^2-1)u^2+{\cal O}(u^3)$ where the subscript indicates that model $\hat{\cal H}_2$ is under scrutiny. The first moment of this distribution evaluated in $u=0$, as requested for the calculation of the average work, gives us $\langle W_{\hat{\cal H}_2}\rangle=0$, at variance with the result for the average work valid for Eq.~\eqref{model}. The reason behind such dissimilarity should be traced back to the energy-conserving nature of model $\hat{\cal H}_2$, which does not give rise to any squeezing of the oscillators. 
 

Let us go back now to the case embodied by Hamiltonian $\hat{\cal H}_1$. The results gathered so far for a two-element system can be generalised to an array of arbitrary length. In particular, the change in free energy for an array of $N$ harmonic oscillators interacting according to the Hooke-like model 
\begin{equation}
\label{modelloN}
\hat{\cal H}_1=\frac{\Omega}{2}\sum^{N}_{j=1}(\hat x^2_j+\hat p^2_{j})+g_t\sum^{N-1}_{j=1}(\hat x_j-\hat x_{j+1})^2
\end{equation}
reads
\begin{equation}
\Delta F_N=-\frac{1}{\beta}\ln\left[\frac{\sinh^N(\beta\omega/2)}{\Pi^N_{j=1}\sinh(\beta\mu_j/2)}\right]
\end{equation}
with $\mu_j=\omega\sqrt{\lambda_j/\omega}$, $\omega=\Omega/2$ and $\{\lambda_j\}$ the set of eigenvalues of the adjacency matrix representing the Hamiltonian $\hat{\cal H}_1$ (cf. the Appendix). Using the characteristic function for coherent states $\chi_{\{\alpha\}}(u)$ given in Eq.~\eqref{tantiN} and its first statistical moment,  
we can easily calculate the average work, which is found to scale with the number of oscillators as 
\begin{equation}
\label{workN}
\langle W\rangle_N=g_0 V\frac{N-1}{2}.
\end{equation}
This formula has a very simple interpretation. Each interaction term (there are in total $N-1$ of them) brings in a contribution $g_0 V/2$ to the total work.  The factor $N-1$ can also be understood by noticing the fact that, out of the $N$ modes involved in the evolution of the system resulting from the quench, only $N-1$ of them are squeezed. This is proven rigorously in the Appendix, where the spectrum of Eq.~\eqref{modelloN} is shown to always contain the bare-oscillator value $\omega$ among $N-1$ squeezing-dependent values [cf. Eq.~\eqref{spettro}]. Physically, this is due to the fact that the centre-of-mass mode of the system of oscillators is always a normal mode of the system itself.

\begin{figure}[t]
\includegraphics[width=0.85\linewidth]{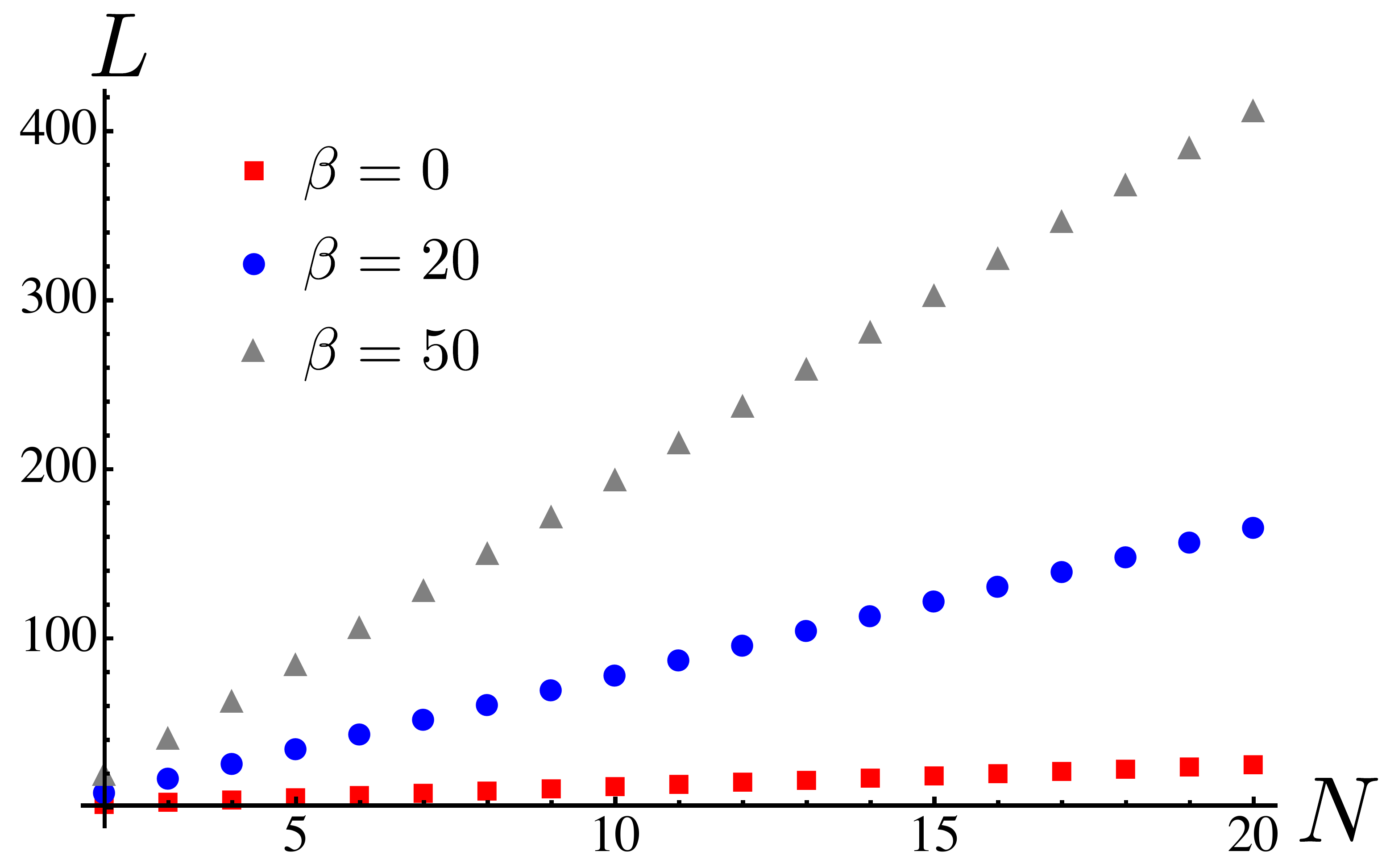}
\caption{(Color online) Nonequilibrium lag after a quantum quench in an array of $N$ Hooke-like coupled harmonic oscillators with $g=2\omega$ and for three values of the inverse temperature $\beta$.} 
\label{entorpycontroN} 
\end{figure}

With the average work and the change in free energy, we can finally consider the nonequilibrium lag  for $N$ oscillators
\begin{equation}
\begin{aligned}
L&
=\frac{\beta g_0 V (N-1)}{2}+\ln\left[\sinh^N\left(\frac{\beta\omega}{2}\right)\right]-\sum^N_{j=1}\ln\left[\sinh\left(\frac{\beta\mu_j}{2}\right)\right]\\
&=(N-1)\left(\frac{\beta g_0V}{2}+\ln\left[\sinh\left(\frac{\beta\omega}{2}\right)\right]\right)-\sum^N_{j=2}\ln\left[\sinh\left(\frac{\beta\mu_j}{2}\right)\right].
\end{aligned}
\end{equation}
The behavior of $L$ against the length of the chain and for three values of the inverse temperature $\beta$ is reported in Fig.~\ref{entorpycontroN}.

\section{Relation with quantum correlations}
\label{correlations}

In the following, we 
study the possibility of establishing a direct quantitative link between the nonequilibrium lag  produced by the quantum quench under scrutiny and the general quantum correlations shared by the oscillators. We will mainly restrict our attention to a two-oscillator system, so as to avoid unnecessary computational problems. 

Fig.~\ref{studio} and our related analysis have shown the existence of a one-to-one correspondence between temperature and the nonequilibrium lag  ${ L}$, which can be considered as a reliable {\it thermometer}, in particular in the interesting quantum region of $\beta\gg1$. In a qualitatively analogous way, it is possible to establish a link between $\beta$ and the amount of non-classical correlations (as measured by Gaussian entanglement and discord) shared by the oscillators of our array after the quench. 

\begin{figure*}[t]
 {\bf (a)}\hskip7cm{\bf (b)}
\includegraphics[width=0.9\columnwidth]{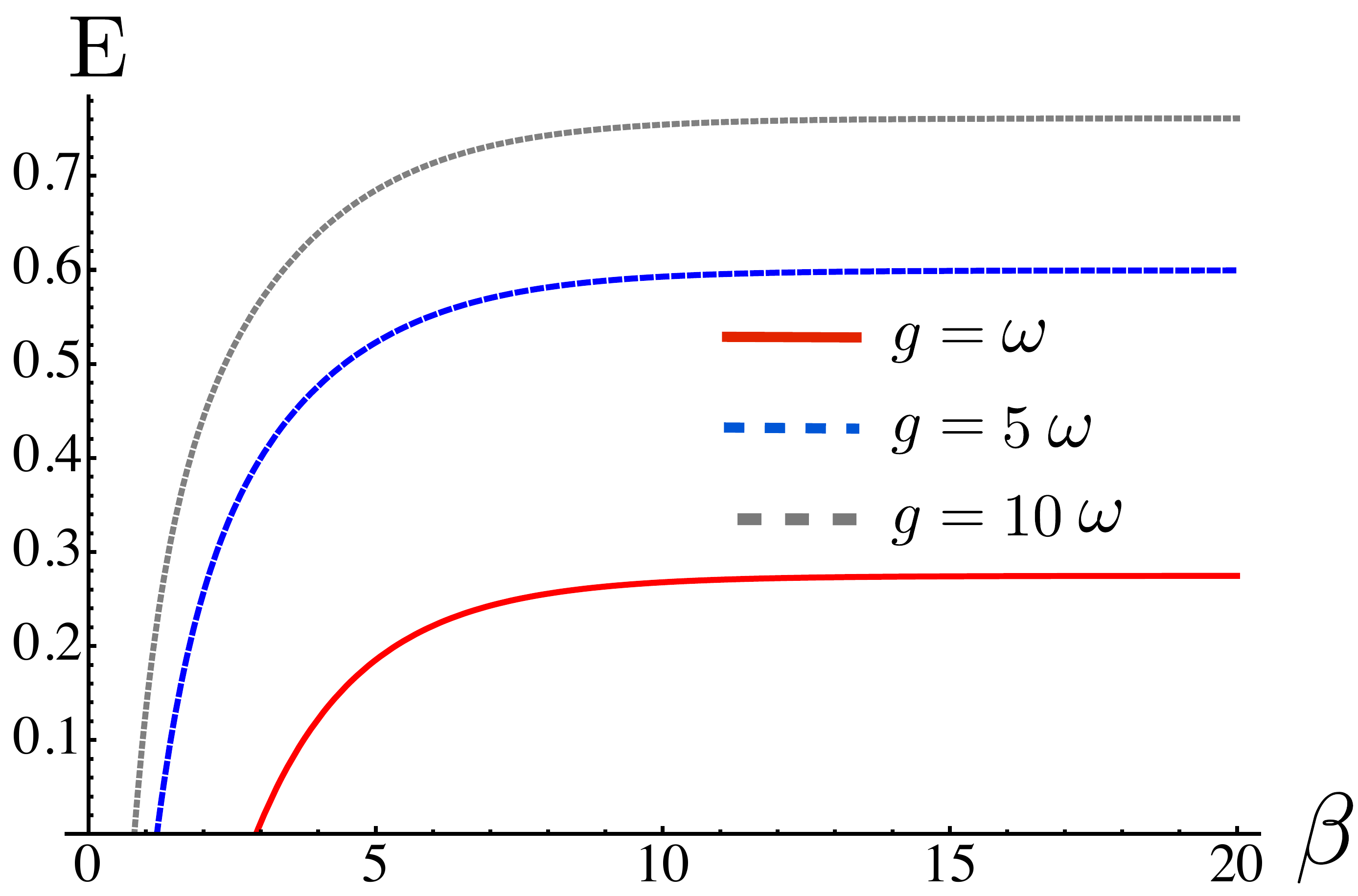}\includegraphics[width=0.95\columnwidth]{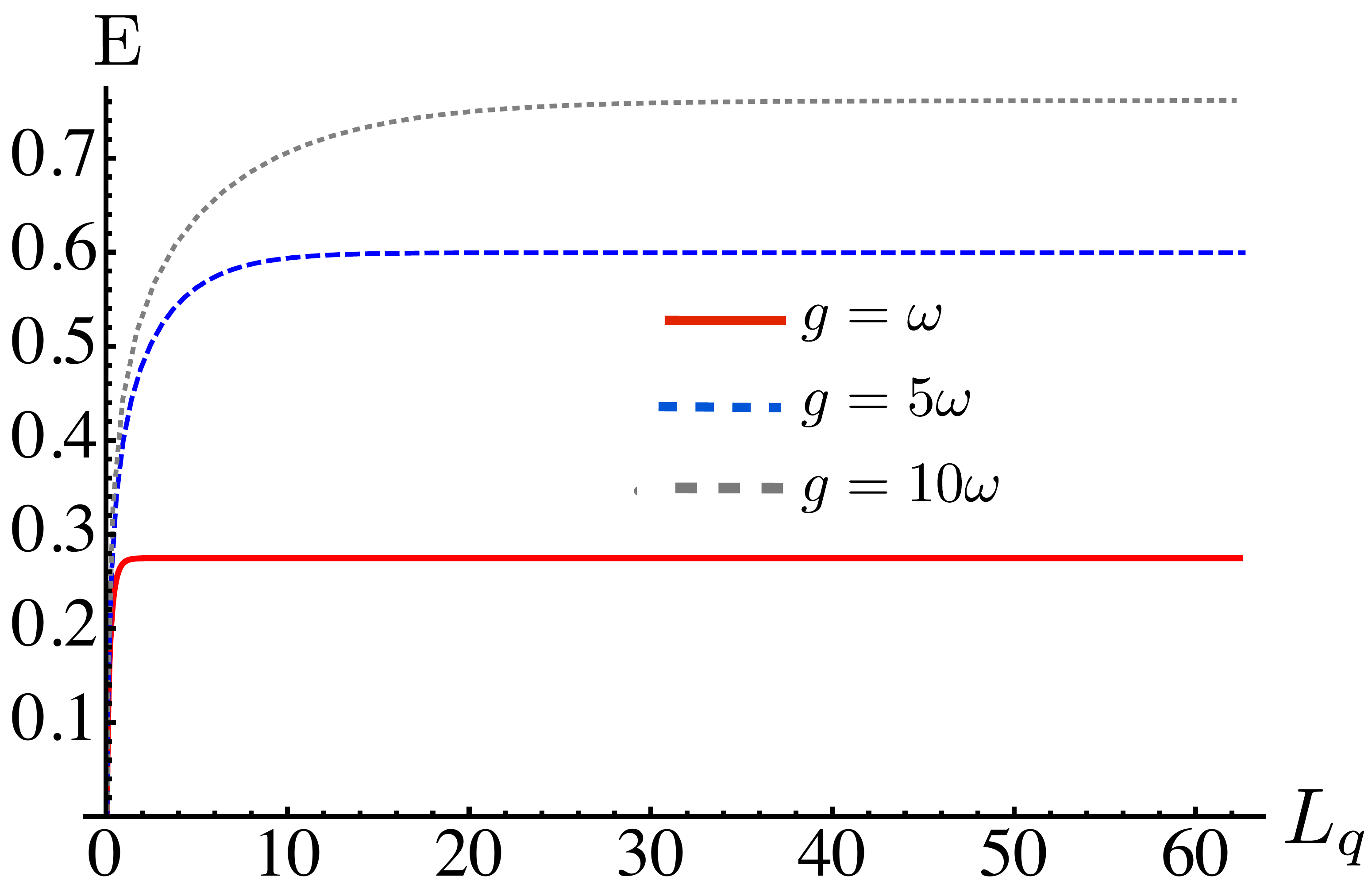}
\caption{(Color online) {\bf (a)} Entanglement in the equilibrium state of two harmonic oscillators coupled by a Hooke-like model, plotted against the inverse temperature $\beta$ for three values of the coupling strength $g$ (values given in units of $\omega$). {\bf (b)} Illustration of the link between $L_q$ and the logarithmic negativity in a system of two Hooke-like coupled harmonic oscillators shown for three different values of the quench amplitude. The inverse temperature $\beta$ is the curvilinear abscissa of each curve.} 
\label{entanglement} 
\end{figure*}

We start by addressing entanglement, which is quantified here using the logarithmic negativity. For a two-mode Gaussian state, such as the one corresponding to the equilibrium state of Hamiltonian in Eq.~\eqref{model} at inverse temperature $\beta$ , the latter is defined as 
\begin{equation}
\text{E}=\max[0,-\ln\nu_-].
\end{equation}
Here, $\nu_-$ is the smallest eigenvalue of the matrix $|i{\bm \Sigma}{\text P}{\bm \sigma}{\text P}|$, where ${\text P}=\text{diag}[1,1,1,-1]$ performs the inversion of momentum of the second harmonic oscillator, ${\bm \Sigma}=i{\bm\sigma}_y\otimes{\bm \sigma}_y$ is the symplectic matrix (with $\sigma_y$ the y-Pauli matrix) and ${\bm \sigma}$ is the covariance matrix of the two-oscillator system~\cite{ferraro}. The latter can be easily calculated using the formal analogy with an optical interferometer discussed above and used to calculate the characteristic function of the work distribution. The results of our calculations are shown in Fig.~\ref{entanglement}, where the logarithmic negativity is plotted against the inverse temperature at three values of the quench amplitude. Analytically
\begin{equation}
\text{E}=\max\left[0,-\ln\frac{\sqrt{\left[1+\text{csch}\left(\frac{\beta  \omega }{2}\right)\right]\left[1+\text{csch}\left(\frac{\beta\omega}{2} 
   \sqrt{1+\frac{2 g_0}{\omega }}\right)\right]}}{\sqrt[4]{\left({1+2 g_0/\omega }\right)}}\right],
\end{equation}
which reaches the maximum value given by $E=\ln\sqrt[4]{1+2g_0/\omega}$ for $\beta\to\infty$. The two-oscillator entanglement disappears above a threshold temperature whose value depends on the ratio $g_0/\omega$. 

\begin{figure}[!t]
\includegraphics[width=\linewidth]{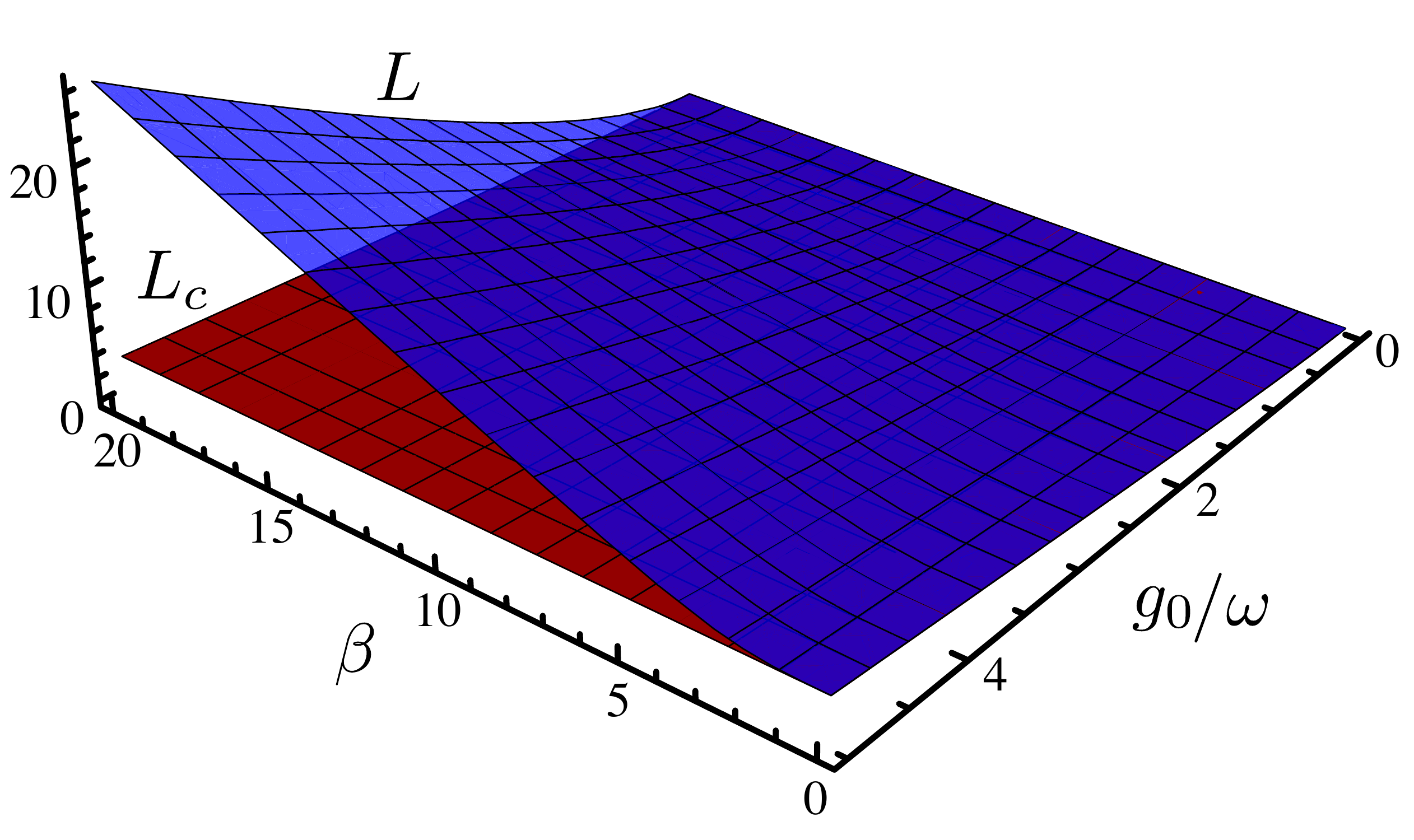}
\caption{(Color online) Comparison between the full form of the nonequilibrium lag $L$ and its classical counterpart $L_c$ shown against the inverse temperature $\beta$ and the dimensionless interaction strength $g_0/\omega$. At high temperature $L\to L_c$, regardless of the strength of the quench.} 
\label{nonequilag} 
\end{figure}

We now aim at comparing the behavior of $\text{E}$ to that of the `quantum' part of the  nonequilibrium lag, i.e. the part of ${ L}$ that remains after subtracting the high-temperature value $L_c\equiv\lim_{\beta\to0}{ L}=g_0/\omega-\ln\sqrt[4]{1+2g_0/\omega}$. As seen in Fig.~\ref{nonequilag}, at low temperatures and large coupling strengths, the quantum part of $L$ is crucial in determining quantitatively the non equilibrium lag. In Fig.~\ref{entanglement} {\bf (b)} we thus plot the logarithmic negativity against the quantum part $L_q\equiv{ L}-L_c$ of the  nonequilibrium lag, by eliminating the inverse temperature, showing that a direct relation exists between such quantities, which appear to be in mutual functional dependence. The (in general) involved non-linear relation of each of them with the inverse temperature prevents us from finding such dependence explicitly. However, some insight can be gathered from the behavior shown in Fig.~\ref{entanglement} {\bf (b)}, such as the existence of a (quench-dependent) threshold above which the logarithmic negativity becomes insensitive to the actual value of $L_q$. As the inverse temperature embodies the curvilinear abscissa of each of the curves displayed in Figs.~\ref{entanglement}, we can identify the region of insensitivity to the nonequilibrium lag  as the low-temperature part of Fig.~\ref{entanglement} {\bf (a)}. 
However, the large-temperature part of  Fig.~\ref{entanglement} {\bf (b)} is somehow misleading: at large temperature, entanglement is strictly null while $L_q$ might well achieve, in general, non-zero values. As the existence of such a temperature-dependent threshold for the non-nullity of entanglement is an expected common feature of entanglement measures,  this induces us to consider entanglement as a somehow unfit figure of merit for a comparison between the behavior of quantum correlations and the nonequilibrium lag  produced across the process. We thus turn our attention to the measure of quantum correlations embodied by the Gaussian discord~\cite{GiordaParis}: for a Gaussian state with covariance matrix ${\bm\sigma}=\begin{pmatrix}{\bm \alpha}_1&{\bm \gamma}\\{\bm\gamma}&{\bm\alpha}_2\end{pmatrix}$, discord is defined as  
\begin{equation}
{\rm D}=f(\sqrt{\det{{\bm\alpha}_2}})-f(\nu_-)-f(\nu_+)+\inf_{{\bm\sigma}_0}f(\sqrt{\det\epsilon}).
\end{equation}
Here, $f(x)=(x+1)/2\ln[(x+1)/2]-(x-1)/2\ln[(x-1)/2]$, $\nu_\pm$ are the symplectic eigenvalues of ${\bm\sigma}$, ${\bm\epsilon}={\bm\alpha}_1-{\bm\gamma}({\bm\alpha}_2+{\bm\sigma}_0)^{-1}{\gamma}^T$ is the Schur complement of ${\bm\alpha}_1$ and ${\bm\sigma}_0$ is the covariance matrix of a single-mode rotated squeezed state.

\begin{figure*}[t]
{\bf (a)}\hskip7cm{\bf (b)}
\includegraphics[width=0.95\columnwidth]{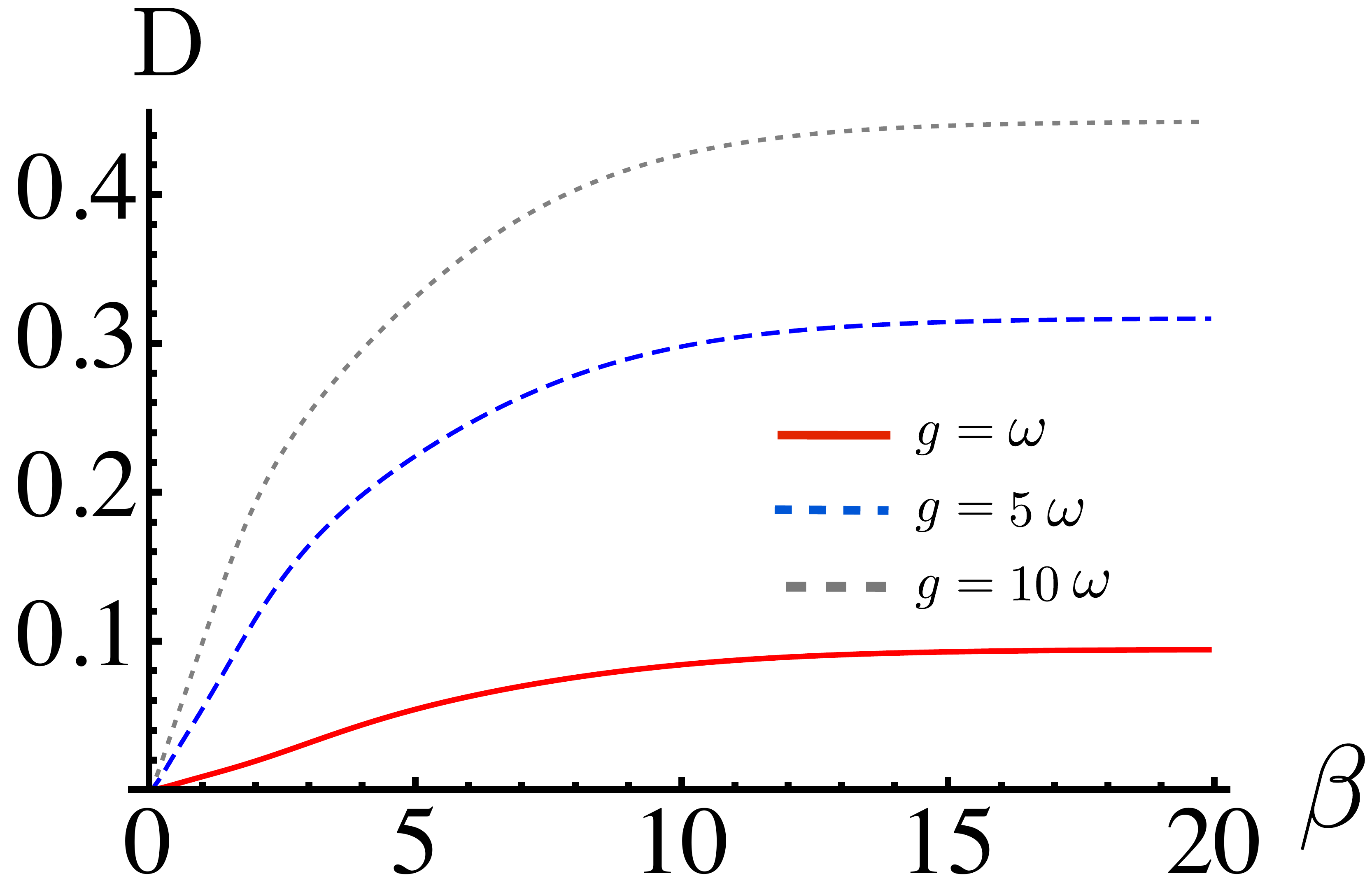}\includegraphics[width=\columnwidth]{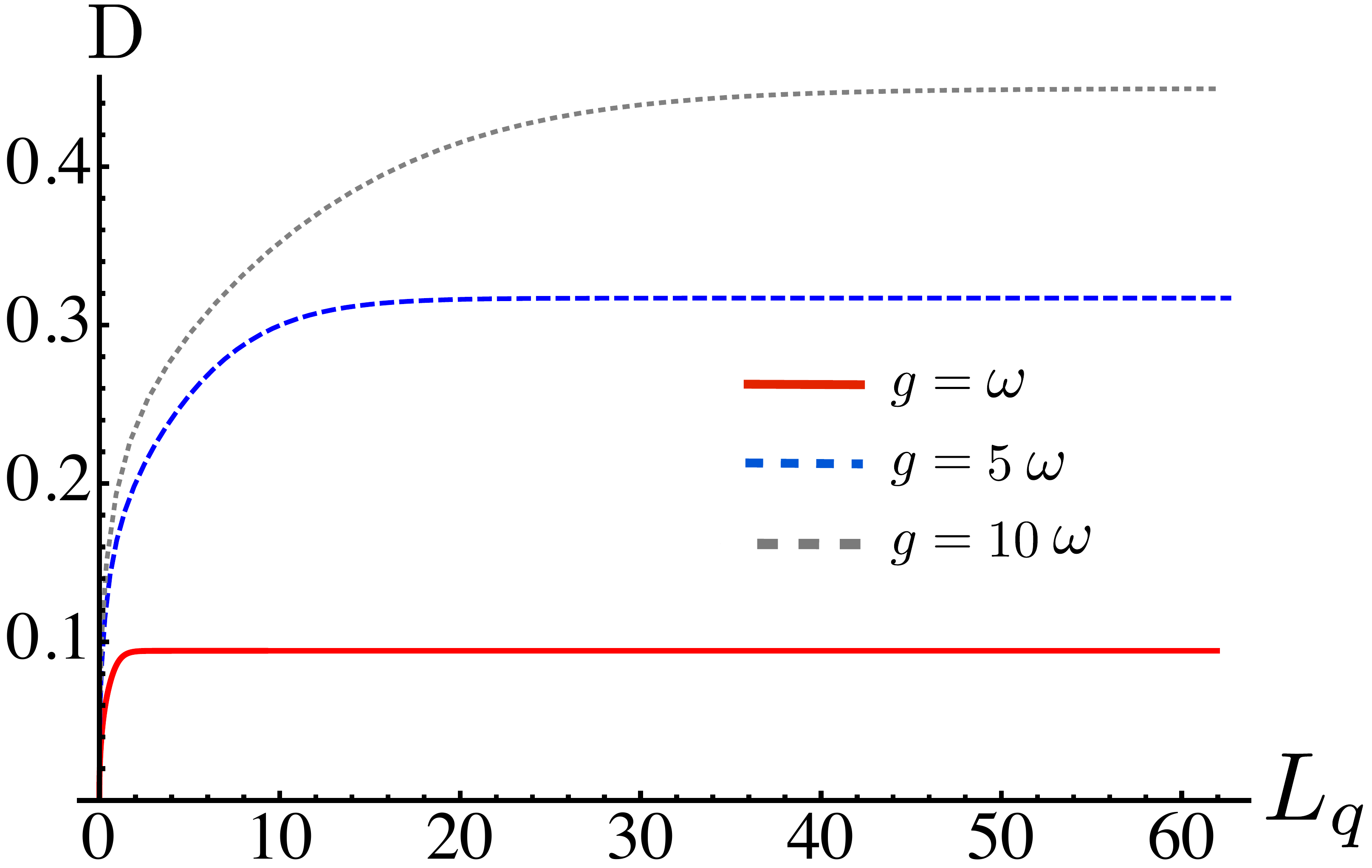}
\caption{(Color online) {\bf (a)} Gaussian discord in the equilibrium state of two harmonic oscillators coupled by a Hooke-like model, plotted against the inverse temperature $\beta$ for three values of the coupling strength $g$ (values given in units of $\omega$). {\bf (b)} Illustration of the link between $L_q$ and the Gaussian discord in a system of two Hooke-like coupled harmonic oscillators shown for three different values of the quench amplitude. The inverse temperature $\beta$ is the curvilinear abscissa of each curve. } 
\label{discord} 
\end{figure*}

The results of the calculations are shown in Fig.~\ref{discord}. First, panel {\bf (a)} shows that, at variance with entanglement, Gaussian discord allows for no threshold in temperature and it disappears only for $\beta=0$. Second, albeit panel {\bf (b)} is qualitatively similar to Fig.~\ref{entanglement} {\bf (b)}, the analysis of the former is less ambiguous as both $D$ and $L_q$ vanish at infinite temperatures only. Although valid for the specific case of our system and so far limited to a study of only two-body quantum correlations, our analysis suggests the existence of a clear functional link between the amount of general quantum correlations established between two of the interacting harmonic oscillators studied here and the amount of nonequilibrium lag  generated in a quantum-quench. It would be interesting to extend our analysis to multipartite figures of merit for quantum correlations. This is, per se, a rather difficult problem due to the current lack of computable quantifiers of genuinely multipartite quantum correlations.

\section{Conclusions} 
\label{conc}

We have characterised the dynamics of relevant quantum and thermodynamic properties of an array of coupled harmonic oscillators in thermal equilibrium and experiencing a sudden quench in the inter-particle coupling strength. We have provided useful analytic expressions for the characteristic function of work distribution, the reversible and dissipated work, and the variation of free energy, which have allowed us to study quantum fluctuation identities in relation to the degree of squeezing induced by the dynamics. Our results showcase an interesting functional dependence of the irreversible lag with respect to the degree of quantum correlations across a two-oscillator system, thus suggesting a direct influence of quantum correlations in the settling of thermodynamic features. 

\section*{APPENDIX}
\label{a1}

We aim at evaluating the function $\chi_{\alpha_1,\alpha_2}(u)={}_{12}\langle{\alpha_1,\alpha_2}\vert e^{iu\hat{\cal H}_f}e^{-iu\hat{\cal H}_i}\ket{\alpha_1,\alpha_2}_{12}$. In what follows, we will use the decomposition of the time-evolution operator in Eq.~\eqref{deco} and the fact that $\exp[-i\hat{\cal H}_iu]=\bigotimes^2_{j=1}e^{i\theta_j(t)(\hat x^2_j+\hat p^2_j)}$. We find 
\begin{widetext}
\begin{equation}
\label{esplicito}
\begin{aligned}
\chi_{\alpha_1,\alpha_2}(u)&=e^{i\frac{\theta_2(u)-\omega u}{2}}{}_1\!\bra{\alpha_-}{}_2\!\bra{\alpha_+}\hat{\cal S}^\dag_2(r)\hat{\cal S}_2(r e^{2i\theta_2(u)})\ket{\alpha_-}_1\vert{\alpha_+e^{-i\theta_1(u)+i\theta_2(u)}}\rangle_2\\
&=e^{i\frac{\theta_2(u)-\omega u}{2}}{}_2\!\bra{0}\hat{\cal D}^\dag_2(\alpha_+)\hat{\cal S}^\dag_2(r)\hat{\cal S}_2(re^{2i\theta_2(u)})\hat{\cal D}(\alpha_+e^{-i\theta_1(u)+i\theta_2(u)})\ket{0}_2.
\end{aligned}
\end{equation}
\end{widetext}
with $\alpha_\pm=(\alpha_1\pm\alpha_2)/\sqrt 2$. Eq.~\eqref{esplicito} can be put into the form of an overlap between displaced squeezed states by exploiting the operator identity
\begin{equation}
\hat{\cal S}(\xi)\hat{\cal D}(\zeta)\hat{\cal S}^\dag(\xi)=\hat{\cal D}(\zeta\cosh|\xi|+\zeta^*e^{i\arg\xi}\sinh|\xi|),
\end{equation}
which is valid for any $\zeta,\xi\in\mathbb{C}$. The order of squeezing and displacement operators can thus be swapped to get $\chi_{\alpha_1,\alpha_2}(u)=e^{i\frac{\theta_2(u)-\omega u}{2}}\!\bra{\zeta_1;\xi_1}\zeta_2;\xi_2\rangle$ with
\begin{equation}
\begin{aligned}
\zeta_1&={\alpha_+}\cosh r+{\alpha^*_+}\sinh r,\\
\zeta_2&=[{\alpha_+}e^{-i\omega u}\cosh r+{\alpha^*_+}e^{i\omega u}\sinh r]e^{i\theta_2(u)},\\
\xi_1&=r,~\xi_2=r e^{2i\theta_2(u)}.
\end{aligned}
\end{equation}

We now sketch the formal procedure for the generalization of the approach discussed above to the case of a harmonic chain of an arbitrary number of oscillators coupled through the Hooke-like model
\begin{equation}
\label{modelN}
\hat{\cal H}^N_1=\omega\sum^N_{j=1}(\hat x^2_j+\hat p^2_j)+g_t\sum^{N-1}_{j=1}(x_j-x_{j+1})^2,
\end{equation}
which generalises Eq.~\eqref{model}. In the basis of the quadratures $\hat{\bm r}=(\hat x_1,\dots,\hat{x}_N,\hat p_1,\dots,\hat  p_N)^T$, the Hamiltonian is represented by the block matrix $\hat{\cal H}^N_1=\hat{\bm r}^{T}{H}^N_1\hat{\bm r}$ reading
\begin{equation}
H^N_1=
\begin{pmatrix}
{\mathbb V}&{\mathbb O}\\
{\mathbb O}&{\mathbb K}
\end{pmatrix}
\end{equation}
with ${\mathbb O}$ the identically null matrix, ${\mathbb K}=\omega\openone_N$ the matrix representing the kinetic-energy term and 
\begin{equation}
\label{V}
{\mathbb V}=
\begin{pmatrix}
\omega+g_t & -g_t & & 0\\
-g_t & \omega+2g_t & -g_t & \\
& \ddots & \ddots &  \\
& -g_t & \omega+2g_t & -g_t \\
0 & & -g_t & \omega+g_t \end{pmatrix}
\end{equation}
\noindent
that stands the potential energy of the Hamiltonian. Eq.~\eqref{V} embodies a symmetric quasi-uniform tridiagonal (QUT) matrix, whose spectrum can be fully characterised analytically. In fact, by shifting and rescaling its entries as $-(1/g_t)[{\mathbb V}-(\omega+2g_t)\openone_N]$, we get a special case of the QUT matrices explicitly addressed in Ref.~\cite{Banchi}. The eigenvalues $\{\lambda_j\}$ of such matrix can be analytically computed and give
\begin{equation}
\label{spettro}
\lambda_j=\omega+2g_t\left(1-\cos\left[\frac{\pi(j-1)}{N}\right]\right)~~~~j=1,..,N
\end{equation}
which shows that there is always one eigenvalue equal to the bare oscillator frequency $\omega$. As we will see, this has quite remarkable consequences and is strongly tied with the results valid for the two-oscillator case addressed in the main text. The diagonalization of ${\mathbb V}$ is achieved through an orthogonal matrix ${\mathbb P}$ (which can be fully determined regardless of $N$~\cite{Banchi}) that leaves ${\mathbb K}$ unaffected. Following the general protocol put forward in Ref.~\cite{Reck}, such matrix can be easily broken down into a cascade of beam-splitters and phase rotators. Therefore
\begin{equation}
PH^N_1P^{T}\equiv H_{D^N_1}=\begin{pmatrix}
{\mathbb V_D}&{\mathbb O}\\
{\mathbb O}&{\mathbb K}
\end{pmatrix}
\end{equation}
with $P={\mathbb P}^{T}\oplus{\mathbb P}$ and ${\mathbb V}_D=\text{diag}[\lambda_1,\dots,\lambda_N]$. Matrix $H_{D^N_1}$ corresponds to a Hamiltonian term of the form
\begin{equation}
\label{Nm1sq}
\hat{\cal H}_{{\cal D}^N_1}=\omega(\hat X^2_1+\hat P^2_1)+\sum^N_{j=2}[\lambda_j\hat X^2_j+\omega\hat P^2_j]
\end{equation}
with $(\hat X_j,\hat P_j)$ the new modes of the system. Eq.~\eqref{Nm1sq} has been deliberately written in a way to emphasize that only $N-1$ oscillators are squeezed. Therefore, by applying the squeezing operator $\hat{\cal S}^N=\openone_1\otimes\left[\otimes^{N}_{j=2}\hat{\cal S}_j(r_j)\right]$ we can transform the time-evolution operato generated by the initial model \eqref{modelN} as
\begin{equation}
\hat{U}^N(t)=e^{-i\hat{\cal H}^N_1 t}=\hat{\cal P}^\dag\hat{\cal S}^{N\dag}\left[\otimes^N_{j=1}\hat{\cal R}_j(\theta_j(t))\right]\hat{\cal S}^N\hat{\cal P}
\end{equation}
with $\hat{\cal P}$ the operator corresponding to the transformation matrix $P$ and $\theta_j(t)=\lambda_j t$. This is in formal correspondence with what has been illustrated for the two-oscillator case. 

\begin{figure*}[t]
{\bf (a)}\hskip8.5cm{\bf (b)}\\
\includegraphics[width=0.7\columnwidth]{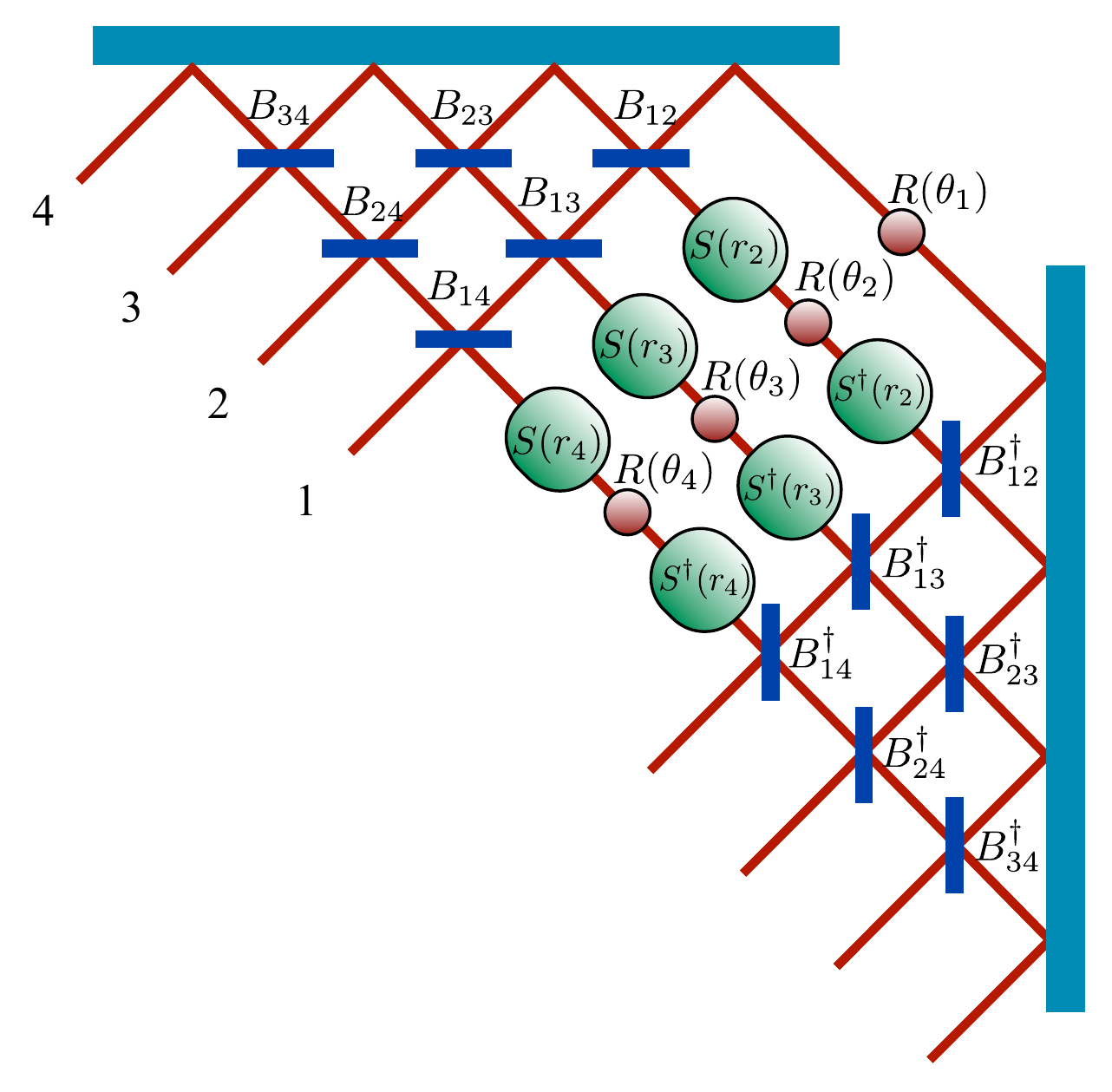}\hskip3.5cm\includegraphics[width=0.7\columnwidth]{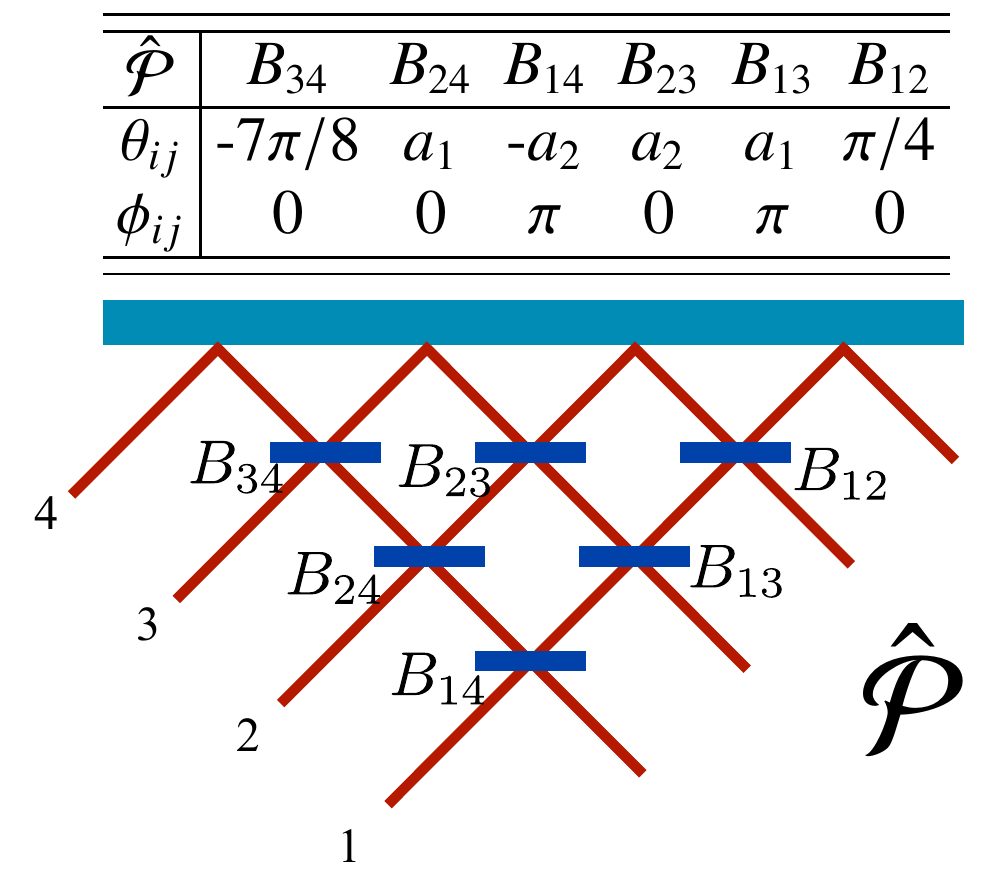}
\caption{(Color online) {\bf (a)} Equivalent interferometer that describes the evolution of a set of four Hooke-like coupled harmonic oscillators. As we have described, such decompostion enables the calculation of the characteristic function of work distribution following a quench of the coupling strength among the oscillators. The configuration and parameters of the array of beam splitters are determined as discussed in the Appendix. The set of squeezing operations $\hat S(r_j)~j=2,..,4$ and phase-space rotations $\hat R(\theta_j)~(j=1,..,4)$ comple the decomposition. {\bf (b)} Linear-optics decomposition of the transformation $\hat{\cal P}$ that diagonalizes the Hamiltonian of $4$ coupled harmonic oscillators. We show the arrangement of beam splitters $B_{ij}$ between modes $i$ and $j$ needed for the decompostion, as well as the corresponding values of the parameters $\theta_{ij}$ and $\phi_{ij}$. In the table, we have set $\tan(a_1)=-({4-\sqrt8})^{1/2}$, and $\tan(a_2)=({7+\sqrt{32}})^{1/2}$. }
\label{completo} 
\end{figure*}

Let us concentrate now on the (so far unspecified) operator $\hat{\cal P}$. As mentioned, this can be decomposed into a suitable sequence of beam-splitting and phase-rotation operations. For the sake of completeness, in Fig.~\ref{completo} {\bf (a)} and {\bf (b)} we provide a pictorial representation of the equivalent interferometer and the sequence of beam-splitting and phase-rotation operations needed for the case of four oscillators. However, although useful in order to identify the correct sequence of operations that would realise $\hat{\cal P}$, we do not actually need to determine the full-fetched decomposition in order to be able to understand the effect that such transformation has overall. Indeed,  it is enough to have the entries of $P$ to determine the transformation laws of the oscillators' quadratures as $\hat r_i\to \sum^N_{j=1}P_{ji}\hat r_j$ $(r=x,p)$. It takes a straightforward calculation to check that, when applied to the tensor product of $N$ coherent states $\otimes^N_{i=1}\ket{\alpha_i}_i$, this leads to 
\begin{equation}
\otimes^N_{i=1}\ket{\alpha_i}_i\to e^{i\varphi(P)}\bigotimes^N_{i=1}\ket{\sum^N_{j=1}P_{ji}\hat \alpha_i}_i,
\end{equation}
with $\varphi(P)$ a phase that depends on the set of amplitudes $\alpha_i$ and the entries of $P$. Therefore, the calculation of the characteristic function of the work distribution for an initial thermal equilibrium state of $N$ coupled harmonic oscillators can proceed along the lines of the approach sketched in the main text for two modes only, resulting in 
\begin{equation}
\chi(u)=\int d^2\alpha_1\cdots\int d^2\alpha_n\Pi^N_{j=1}P^{th}_{V}(\alpha_j)\chi_{\{\alpha\}}(u)
\end{equation}
with $\chi_{\{\alpha\}}(u)$ the characteristic function of work for a collection of $N$ modes, each initially prepared in a coherent states of amplitude $\alpha_j$ and reading 
\begin{equation}
\label{tantiN}
\chi_{\{\alpha\}}(u)=e^{\frac{i}{2}\sum^N_{j=1}\theta_j(u)-i\frac{N}{2}\omega u}\Pi^N_{j=2}{}\langle\zeta_{1,j};\xi_{1,j}|\zeta_{2,j};\xi_{2,j}\rangle.
\end{equation}
Here,  $\zeta_{1(2),j}$ and $\xi_{1(2),j}$ are the amplitudes of the displacement and squeezing operations, respectively, of the displaced squeezed states of mode $j=2,..,N$ that enter into the definition of $\chi_{\{\alpha\}}(u)$. Their expressions can be gathered easily in a way analogues to what has been done for just two oscillators.
 
\acknowledgments 
MP thanks Leonardo Banchi for useful discussions on the topic of Ref.~\cite{Banchi}. AC acknowledges the Northern Ireland DEL for support. LM is supported by the EU through a Marie Curie IEF Fellowship. 
MP acknowledges hospitality by the Centro de Ci\^encias Naturais e Humanas at the Universidade Federal do ABC (UFABC) during the early stages of this work. 
This work has been supported by the UK EPSRC (EP/G004579/1 and EP/L005026/1), the Alexander von Humboldt Stiftung, the John Templeton Foundation (grant ID 43467), and the EU Collaborative Project TherMiQ (Grant Agreement 618074). MC thanks the Volkswagen Foundation (project No. I/83902). FLS is a member of the Brazilian National Institute of Science and Technology of Quantum Information (INCT-IQ) and acknowledges partial support from CNPq (grant nr. 308948/2011-4). FLS and MP are supported by the CNPq ``Ci\^{e}ncia sen Fronteiras'' programme through the ``Pesquisador Visitante Especial'' initiative (grant nr. 401265/2012-9). 
VV acknowledges funding from the National Research Foundation (Singapore), the Ministry of Education (Singapore), the EPSRC (UK), the Templeton Foundation, the Leverhulme Trust, the Oxford Martin School and the Fell Fund (Oxford).


\begin{thebibliography}{99}


\bibitem{esposito}
M. Esposito, U. Harbola, and S. Mukamel,
Rev. Mod. Phys.Ê {\bf 81}Ê 1665--1702Ê (2009).

\bibitem{campisi} M.~Campisi, P.~H\"{a}nggi, and P.~Talkner, Rev.~Mod.~Phys. {\bf 83}, 771 (2011), {\emph ibid. }
 {\bf 83}, 1653 (2011).

\bibitem{seifert} U. Seifert, Rep. Prog. Phys. {\bf 75}, 126001 (2012).

\bibitem{Tasaki} H.~Tasaki,  {arXiv:cond-mat/0009244v2};  J.~Kurchan, {arXiv:cond-mat/0007360v2}; S. Mukamel,  Phys. Rev. Lett. {\bf 90}, 170604 (2003).

\bibitem{Crooks}G.~E.~Crooks, Phys.~Rev.~E {\bf 60}, 2721 (1999).

\bibitem{Jarzynski} C.~Jarzynski, Phys.~Rev.~Lett. {\bf 78}, 2690 (1997).

\bibitem{lutz} P.~Talkner, E.~Lutz and P.~H\"anggi, Phys.~Rev.~E {\bf 75}, 050102R (2007).

\bibitem{varie}  J. P. Pekola, P. Solinas, A. Shnirman, and D. V. Averin, arXiv:1212.5808 (2012); V. Vedral, arXiv:1204.6168 (2012); J. Phys. A: Math. Theor. {\bf 45}, 272001 (2012); K. Micadei, R. M. Serra, L. C. Celeri, arXiv:1211.0506 (2012); D. Kafri and S. Deffner, Phys. Rev. A {\bf 86}, 044302 (2012).

\bibitem{Abah} O. Abah, J. Rossnagel, G. Jacob, S. Deffner, F. Schmidt-Kaler, K. Singer, and Eric Lutz, Phys. Rev. Lett {\bf 109}, 203006 (2012).

\bibitem{Dorner} R. Dorner, J. Goold, C. Cormick, M. Paternostro, and V. Vedral, Phys. Rev. Lett. {\bf 109}, 160601 (2012).

\bibitem{Joshi} A. Silva, Phys. Rev. Lett. {\bf 101},120603 (2008). 

\bibitem{Huber} G. Huber, F. Schmidt-Kaler, S. Deffner and E. Lutz, Phys. Rev. Lett. {\bf 101}, 070403 (2008).

\bibitem{Heyl} M. Heyl, and S. Kehrein, Phys Rev Lett {\bf 108}, 190601 (2012).

\bibitem{Ngo} V. A. Ngo, and S. Haas, Phys. Rev. E {\bf 86}, 031127 (2012); T. Albash, D. A. Lidar, M. Marvian, and P. Zanardi, {\it ibid.} {\bf 88}, 032146 (2013).

\bibitem{CPmaps} M. Campisi, P. Talkner, and P. H\"anggi, Phys. Rev. Lett. {\bf 102}, 210401 (2009); 

\bibitem{Oxf} R. Dorner, S. R. Clark, L. Heaney, R. Fazio, J. Goold and V. Vedral, Phys. Rev. Lett. {\bf 110}, 230601 (2013); L. Mazzola, G. De Chiara, and M. Paternostro, Phys. Rev. Lett. {\bf 110}, 230602 (2013); L. Mazzola, G. De Chiara, and M. Paternostro, arXiv:1401.0566 (2014).

\bibitem{KavanJohn} J. Goold, and K. Modi, arXiv:1401.4088 (2014).

\bibitem{JohnMauroKavan} J. Goold, M. Paternostro, and K. Modi, arXiv:1402.4499 (2014).

\bibitem{Batalhao} T. B. Batalh\~ao, A. M. Souza, L. Mazzola, R. Auccaise, I. S. Oliveira, J. Goold, G. De Chiara, M. Paternostro, and R. M. Serra, arXiv:1308.3241 (2013).


\bibitem{Smacchia} P. Smacchia, and A. Silva, Phys. Rev. E {\bf 88}, 042109 (2013).

\bibitem{Joshi13EPJB86} D. G. Joshi, M.~Campisi, The European Physical Journal B \textbf{86}, 157 (2013).

\bibitem{Sindona} A. Sindona, N. Lo Gullo, J. Goold, and F. Plastina, arXiv:1309.2669 (2013)

\bibitem{dudu} E. Mascarenhas, H. Bragan$\c{c}$a, R. Dorner, M. Fran$\c{c}$a Santos, V. Vedral, K. Modi, and J. Goold, arXiv:1307.5544 (2013).

\bibitem{Fusco} L. Fusco, {\it et al.}, (to appear, 2014).

\bibitem{varieoscillator} S. Deffner, and E. Lutz, Phys. Rev. E {\bf 77}, 021128 (2008); M. Campisi, {\it ibid.} {\bf 78}, 051123 (2008); P. Talkner, P. Sekhar Burada, and P. H\"anggi, {\it ibid.} {\bf 78}, 011115 (2008); T. Monnai, Phys. Rev. E {\bf 81}, 011129 (2010); S. Deffner, O. Abah, and E. Lutz, Chem. Phys. {\bf 375}, 200 (2010); J. M. Horowitz, Phys. Rev. E {\bf 85}, 031110 (2012).

\bibitem{Galve} F. Galve, and E. Lutz, Phys. Rev. A {\bf 79}, 055804 (2009).

 

\bibitem{Caves} C. M. Caves, Phys. Rev. D {\bf 23}, 1693 (1981).

\bibitem{Moeller} K. B. M\o ller, T. G. J\o rgensen, and J. P. Dahl, Phys. Rev. A {\bf 54}, 5378 (1996).

\bibitem{Helen} M. Paternostro, H. McAneney, and M. S. Kim, Phys. Rev. Lett. {\bf 94}, 070501 (2005).

\bibitem{Banchi} L. Banchi, and R. Vaia, J. Math. Phys. {\bf 54}, 043501 (2013).

\bibitem{Reck} M. Reck, A. Zeilinger, H. J. Bernstein, and P. Bertani, Phys. Rev. Lett. {\bf 73}, 58 (1994).

\bibitem{ferraro} A. Ferraro, S. Olivares, and M. G. A. Paris, {\it Gaussian states in continuous variable quantum information} (Bibliopolis, Napoli, 2005).

\bibitem{GiordaParis} P. Giorda and M. G. A. Paris, Phys. Rev. Lett. {\bf 105}, 020503 (2010).

\bibitem{Bochkov81aPHYSA106}
G.N. {Bochkov}, Y.E. {Kuzovlev}, Physica A \textbf{106}, 443 (1981).

\bibitem{Schloegl66ZP191}
F.~Schl{\"o}gl, Z. Phys. \textbf{191}, 81 (1966).

\bibitem{Vaikuntanathan09EPL87}
S.~Vaikuntanathan, C.~Jarzynski, EPL \textbf{87}, 60005 (2009).

\bibitem{Deffner10PRL105}
S.~Deffner, E.~Lutz, Phys. Rev. Lett. \textbf{105}, 170402 (2010).

\bibitem{Michele} M. Campisi, Stud. Hist. Phil. Mod. Phys.Ê {\bf 39}, 181 (2008).

\bibitem{GibbsBook}
J.~Gibbs, \emph{Elementary Principles in Statistical Mechanics} (Yale U. P.,
  New Haven, 1902)

\bibitem{Hertz10AP338a}
P.~Hertz, Ann. Phys. (Leipzig) \textbf{338}, 225 (1910).

\bibitem{Einstein11AP34}
A.~Einstein, Annalen der Physik \textbf{34}, 175 (1911).

\bibitem{BeckerBook}
R.~Becker, \emph{Theory of Heat} (Springer, New York, 1967)

\bibitem{Campisi08PRE78b}
M.~Campisi, Phys. Rev. E \textbf{78}, 051123 (2008).

\bibitem{MuensterBook}
A.~M\"unster, \emph{Statistical thermodynamics}, Vol.~1 (Springer, Berlin,
  1969)

\bibitem{Campisi05SHPMP36}
M.~Campisi, Stud. Hist. Phil. Mod. Phys. \textbf{36}, 275 (2005).

\bibitem{Dunkel14NATPHYS10}
J.~Dunkel, S.~Hilbert, Nat Phys \textbf{10}, 67 (2014).


\bibitem{tocome} A. Carlisle, {\it et al.}, to appear (2014).

\end{thebibliography}
\end{document}